\documentclass[12pt]{iopart}
\usepackage{iopams}
\usepackage{amssymb}
\usepackage{setstack}
\usepackage{graphicx}
\usepackage{dcolumn}
\usepackage{bm}
\usepackage{dsfont}
\usepackage{amssymb}
\usepackage[active]{srcltx}

\usepackage{multirow}

\usepackage{soul}

\usepackage{color}

\definecolor{yellow}{rgb}{0.5,0.5,0}

\begin{document}
\date{\today}
\title[Motor proteins traffic regulation by supply-demand balance]{Motor
proteins traffic regulation by supply-demand balance of resources}
\author{Luca Ciandrini$^{1,2, \dagger}$,  Izaak Neri$^{2,3, \dagger}$,
Jean-Charles
Walter$^{2, \dagger}$, Olivier Dauloudet$^{1,2}$, Andrea Parmeggiani$^{1,2}$}
\address{$^{1}$ DIMNP UMR
5235 \& CNRS, Universit\'e Montpellier 2, F-34095, Montpellier, France}
\address{$^{2}$ Laboratoire Charles Coulomb UMR
5221 \& CNRS, Universit\'e Montpellier 2, F-34095, Montpellier, France}
\address{$^{3}$ Max Planck Institute for the Physics of Complex Systems,
Dresden, Germany}
\address{$^\dagger$These authors contributed equally to this work.}
\ead{luca.ciandrini@univ-montp2.fr, izaakneri@gmail.com,
olivier.dauloudet@univ-montp2.fr, jean-charles.walter@univ-montp2.fr,
andrea.parmeggiani@univ-montp2.fr}
\begin{abstract}
In cells and {\it in vitro} assays the number of motor proteins involved in
biological transport processes is far from being unlimited. The cytoskeletal
binding sites are in contact with the same finite reservoir of motors (either
the cytosol or the flow chamber) and hence compete for recruiting the available
motors, potentially depleting the reservoir and affecting cytoskeletal
transport.  
In this work we provide a theoretical framework to study, analytically and
numerically, how  motor density profiles and
crowding along cytoskeletal filaments depend on the competition of motors for
their binding sites. We propose two models in which finite processive motor
proteins actively advance along cytoskeletal filaments and are continuously
exchanged with the motor pool. We first look at homogeneous reservoirs and then
examine the effects of free motor diffusion in the surrounding medium.
We consider as a reference situation  recent {\it in vitro} experimental
setups of kinesin-8 motors binding and moving along microtubule filaments in a
flow chamber. We investigate how the crowding of linear motor proteins moving on
a filament can be regulated by the balance between supply (concentration of
motor proteins in the flow chamber) and demand (total number of polymerised
tubulin heterodimers).  
We present analytical results for the density profiles of bound motors, the
reservoir depletion, and propose novel phase diagrams that present the
formation of jams of motor proteins on
 the filament as a function of two tuneable experimental parameters: the motor
protein concentration and the concentration of tubulins polymerized into
cytoskeletal filaments.  
Extensive numerical simulations corroborate the analytical results for
parameters in the experimental range and 
also address the effects of diffusion of motor proteins in the reservoir.
We then propose experiments to validate our models and discuss how the
"supply-demand" effects can regulate motor
traffic also in {\it in vivo} conditions.
\end{abstract}
\noindent{\it Keywords\/}: Molecular motors, Finite resources, Exclusion
processes, Stochastic modeling

\section{Introduction}
Linear motor proteins are ATPase-driven
machines that bind to cytoskeletal polar filaments and perform linear directed
motion along them~\cite{Alberts,HowardBook, SchliwaM}.  They are classified into
three superfamilies, kinesins, dyneins and
myosins~\cite{schliwa_molecular_2003}.   
These motor proteins regulate many intracellular processes that are necessary
to create and maintain the high level of
organization inside the cell, in particular active transport and force
production.  Active transport driven by motor proteins indeed is essential to
carry cargoes
(i.e.~proteins, organelles, metabolites, etc.) over long enough distances when
diffusion becomes ineffective, e.g. in eukaryotic cilia~\cite{RosenBaum,
Scholey} and axons~\cite{Baas, Hollenbeck, Setou}. Motor proteins
control the transport of organelles in various intracellular processes such as
endo- and exocytosis, and are therefore major actors in membrane traffic
phenomena~\cite{Alberts}. Viruses too can take advantage of directed transport
by binding motor proteins to reach the cell core~\cite{Leop}. Moreover, motor
proteins are involved into other processes not directly related to
cargo transport~\cite{muresan_unconventional_2012}, for example the length
regulation of cytoskeletal filaments~\cite{VargaNat, VargaCell}.
Recent experimental studies \cite{VargaNat, VargaCell, Varg2} have focused on
Kip3, the highly processive kinesin-8 motor of budding yeast moving along
microtubule filaments.   In these {\it in vitro} experiments Kip3 diffuses in a
solution and
binds to the microtubules attached to the slide of the flow chamber
\cite{VargaNat, VargaCell, Varg2}.  
By means of fluorophore labeling the authors reconstructed the
steady-state density profiles of motors on the microtubules,  and
found clear evidence of density shocks whereby a queue of motors accumulates
from
the end tip of the filament. These findings have been interpreted within the
framework of a previously developed model: the totally asymmetric simple
exclusion process with Langmuir kinetics
(TASEP-LK$_{\infty}$)~\cite{Par03,parmeggiani2}. This model exhibits {\it
bulk-localized} shocks of particles that have been called domain walls.  These
kind of density discontinuities are the
result of  exclusion  interactions between particles, 
attachment/detachment dynamics of  particles on the filament, and the presence
of a flow bottleneck at the  end tip.  
Such a model allows us to quantitatively characterize the motor density profiles
on the
filament in terms of microscopic parameters and determine the physical
conditions under
which traffic jams arise.  The latter is also of interest to biological
questions, for instance, in understanding whether the  conditions {\it in
vivo} are such that the formation of jams is avoided \cite{Varg2}.

The {\it in vitro} experiments~\cite{VargaCell, Varg2} are  performed at high
motor protein concentrations and low concentrations of tubulin dimers.  In this
way,  
the reservoir of motors can be considered as unlimited and no depletion of
motors diffusing in the chamber is observed. 
Therefore, the modelling approach introduced with the
TASEP-LK$_{\infty}$~\cite{Par03,parmeggiani2} is a good approximation of the
system studied in~\cite{Varg2}, and it provides testable  predictions of
the corresponding experiments.
However, at lower motor protein concentrations, or higher concentrations of
tubulins, the attachment kinetics  leads to
depletion of motors that could severely affect the overall exchange of 
of motors  between the
filament and
the reservoir.

Here we extend motor protein transport models studied in the literature (such as
the TASEP-LK$_{\infty}$) by considering the limited availability of resources in
the reservoir.   Using these models we thus develop a theoretical framework that
allows us to address the reservoir depletion in {\it in vitro} experiments
performed at motor protein concentrations that are low with respect to the
concentrations of  polymerized tubulin heterodimers. In this regime of limited
resources the results of our modelling framework are expected to deviate
qualitatively from TASEP-LK$_{\infty}$.

We study the reservoir depletion effects on motor protein transport using two
different models. The first model is a totally asymmetric simple exclusion
process in contact with a {\it finite} homogeneous reservoir composed of a
limited number of particles. As it will be justified later, we call this model
TASEP-LK$_{m}$. The TASEP-LK$_m$ adds the concentration of binding sites
(i.e.~the total concentration of tubulin dimers available for the binding
kinesins) as
a new control parameter to the TASEP-LK$_\infty$ model. The second model that we
analyze consists in an exclusion process in contact with a diffusive reservoir
in which unbound particles freely diffuse at a finite rate (similar to the
models defined, e.g., in~\cite{Klumppx,Klumppxx, reviewKlumpp, Muller,
parmeggiani_non-equilibrium_2009, Tsek2008}).
Note that the influence of a limited reservoir on the transport of molecular
motors have already been studied for ribosomes~\cite{Adams, Cookx, CianFinite}.
There is however an important difference between the dynamics of motor proteins
and ribosomes: while the former can detach and attach at any filament binding
site, the latter only attach/detach at the first/last site of the filament,
making the mathematical treatment less elaborate. 

Our results allow us to study the competition between the filaments binding
sites for the number of proteins in the reservoir, and investigate its impact on
transport phenomena driven by motor proteins.
We address analytically and numerically how this competition for finite
resources influences the density profile of motor proteins and the formation of
jams on the filament. 
In particular, we describe how the density profile depends on the ratio between
the concentration of motor proteins and the cytoskeletal binding site
concentrations. Our results could be tested in {\it in vitro} and {\it in vivo}
experiments of molecular motors moving along biofilaments such
as the ones presented in~\cite{Varg2, Seitz, blasius_recycling_2013}, by
administering motor protein
concentrations that are lower or comparable to the concentration of tubulin
dimers.   

In section 2 we define the models we use to describe motor protein transport
at limited availability of resources.  
In section 3 we present the transport equations for TASEP-LK$_{m}$ in the
continuum limit of a mean field approximation. We analytically solve the
equations and derive the expressions for the density profiles of bound motors,
determine the parameter regimes for which we expect jam formation, and obtain
the dependence of the jam length on the supply and demand of resources. In
section 4 we present the outcomes of numerical simulations that explicitly
consider the diffusion of particles in the reservoir, and we obtain the density
profiles of motor proteins bound to the filament. 
We show that TASEP-LK in a diffusive reservoir exhibits a phenomenology similar
to TASEP-LK$_m$. The results of active transport in a diffusive reservoir can
indeed be  interpreted in terms of the analytical  approach derived for
TASEP-LK$_m$. We conclude this paper with a discussion of the obtained results,
and opening perspectives in the context of current experiments and theory.

\section{Models, definitions and notations}
In {\it in vitro} experiments such as~\cite{Varg2, Seitz}, the number of motor
proteins bound to a microtubule filament is regulated by the concentration of
motor proteins
(kinesins)  in the solution of the flow chamber. When motor
proteins attach to the filament their concentration in the reservoir is then
expected to decrease; this induces a concentration depletion in the reservoir
due to the finite number of motor proteins (resources).
The ratio between the total concentration of motor proteins $c$ and the total
concentration of filament binding sites $\mu$, denoted by $m\equiv c/\mu$,
quantifies the absolute availability of resources and determines the motor
protein depletion effect. 
This effect becomes relevant when $m\sim \mathcal{O}(1)$, while its importance
vanishes when the ratio between supply and demand becomes infinitely large, i.e.
$m\rightarrow \infty$.  
\begin{figure}[t]
\begin{center}
\includegraphics[width =1\textwidth]{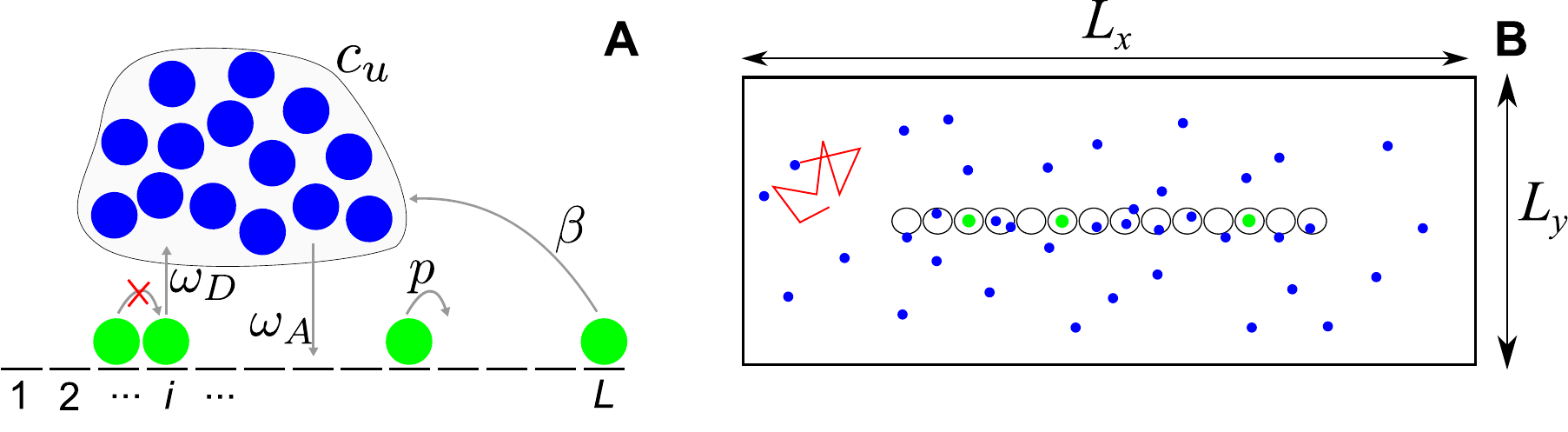}
\caption{The two models studied in this work. \textit{Bound}
particles (in green) proceed along a filament with rate $p$, provided that the
following site of the unidimensional lattice is empty; they are injected with
rate $\omega_{\rm A}=\tilde\omega_{\rm A} c_u$ from a reservoir of
\textit{unbound} particles (in blue) with concentration $c_u$. Particles can
then detach with rate $\omega_D$ on any site, and with rate $\beta$ on the last
site.  The total number of particles $N$ in the combined system of reservoir
and filament is conserved.  In panel (A) we depict the TASEP-LK$_m$ model, with
a
spatially  unstructured
(i.e., homogenous) reservoir, while in the panel (B) we show a spatially
extended
reservoir of dimension $L_x\times L_y$ with closed boundary conditions. We
explicitly consider the free
diffusion of particles in it (a schematic trajectory of an individual particle
is drawn in
red). The sites of the filament have a finite extension modelled by
 disks playing the role of reaction volumes where particles can attach or
detach. Note that a particle can diffuse
 within a reaction volume without attaching.}\label{fig::models}
\end{center}
\end{figure}

To study the depletion effects we define here  two distinct models as sketched
in
figure~\ref{fig::models}: the first model considers an exclusion process in a
finite \textit{homogeneous} reservoir while the second model considers an
exclusion process in a \textit{diffusive} reservoir. The first model is an
extension of the works~\cite{Par03,parmeggiani2, evans_shock_2003} for
homogeneous finite reservoirs, while the second one bears many similarities with
the models with diffusive reservoirs as studied in,
e.g.,~\cite{Klumppx,Klumppxx, reviewKlumpp, Muller,
parmeggiani_non-equilibrium_2009, Tsek2008}. 

\subsection{Active motion of particles on the filament}\label{sub::2.1}
One common way to address the formation of processive motor jams on filaments
is to consider unidimensional driven lattice gases with
steric interactions, namely exclusion processes~\cite{Chou2011, Bresslof,
Chowd}.

The active motion of particles along the filament, as well as their exchange
with the surrounding reservoir, is here formulated as a unidirectional exclusion
process with Langmuir kinetics~\cite{Par03,parmeggiani2, evans_shock_2003}. We
in fact consider that motor protein transport is a continuous time Markovian
hopping process of particles moving unidirectionally along a one-dimensional
lattice at rate $p$, see figure~\ref{fig::models} (A).  
The lattice  has size $L$ with its sites labeled by $i=1..L$. The particles
obey exclusion interactions implying that only one particle can occupy a  single
lattice site. 
Particles represent, for instance, the kinesin-8 motors, and their binding sites
(the lattice sites) correspond to the tubulin dimers forming the microtubule.
Just as kinesins  on cytoskeletal filaments, particles in our models hop
uni-directionally
on the filament.  Particles also attach to and detach from filament binding
sites at, respectively, rates $\omega_A$ and $\omega_D$.  The
attachment-detachment process is therefore described by a Langmuir
adsorption model \cite{Langmuir18}. Both directed transport
and
Langmuir kinetics respect the particle exclusion on the filament. We
furthermore consider that particles reaching the last site of the lattice can
detach from it at a rate $\beta\neq \omega_D$. The different rates and notations
of the model are also defined in figure~\ref{fig::models}.

A meaningful continuum limit keeps $\Omega_A\equiv\omega_AL/p$ and $\Omega_D
\equiv \omega_D
L/p \sim \mathcal{O}(1)$ constant for $L\rightarrow
\infty$~\cite{Par03,parmeggiani2}. The latter scaling implies
that particles cross on average a finite fraction of the filament before
detaching. The fraction between the typical runlength $\ell$ of a single motor
and the filament
length $L\:a$ is $\Omega^{-1}_D$,  with $a$ the site length.  The
quantity 
$\Omega_D$
is therefore a directly measurable
parameter. The attachment rate $\omega_A(i)$ at the filament binding site $i$
depends on
the concentration $c_u(i)$ of motors in the reservoir around the site
$i$, i.e.~$\omega_A(i) = \tilde{\omega}_A c_u(i)$,
where the binding rate constant $\tilde{\omega}_A$ is a parameter to be
determined experimentally.   We further consider the dimensionless
constants $K^{\infty} \equiv c\,\tilde{\omega}_A/\omega_D$ and $\kappa^\infty
\equiv \mu\, \tilde{\omega}_A/\omega_D$ that are, respectively, the renormalized
total concentrations of motor proteins and binding sites (e.g. the concentration
of polymerized tubulin dimers). The finite resource parameter $m$ can then be
written in
terms of these concentrations as $m = K^{\infty}/\kappa^\infty$. We also 
define the Langmuir equilibrium density as $\rho^\infty_L \equiv
K^\infty/(K^\infty+1)$ (a.k.a.~the Langmuir adsorption
isotherm \cite{Langmuir18}), and
the rate constant $\Omega^\infty_A \equiv c\,\tilde{\omega}_A L/p = K^\infty
\Omega_D$ that corresponds to $\Omega_A$ in the case
of unlimited resources.  

We now characterize the features of the reservoir and the  dynamics 
of unbound motors in the two models.

\subsection{Exclusion process in contact with a homogeneous finite reservoir
(TASEP-LK$_m$)}
\label{sec::22}
In the TASEP-LK$_m$ we consider a homogeneous concentration that equals the
total concentration of unbound motors  in the reservoir, i.e.
$c_u(i)=c_u$.
The reservoir adds no additional dynamical rules to the system and we can study
the impact of motor protein depletion on the transport characteristics by using
a simple one-dimensional lattice model. 

In the TASEP-LK$_m$ the motor reservoir is modelled as a box with a number $N_u$
of
unbound particles. Every time a particle detaches from the filament, $N_u$
increases by one unit, while every time a particle from the box attaches to the
filament, $N_u$ decreases by one unit. The concentration of particles in the
reservoir is then given by $c_u = N_u/V$, with $V$ the volume of the reservoir,
see figure~\ref{fig::models} (A). In this way the number of particles in the box
determines the attachment rate on the lattice, and the dynamics on the lattice
is thus coupled with the number of particles in the box. 

Since the total number of motor proteins $N$ is limited (i.e.~$N_u$ is finite),
the rescaled concentration of unbound (or free) motors in the reservoir $K
\equiv c_u\,\tilde{\omega}_{\rm A}/\omega_{\rm D}$ will in general be different
to the total rescaled concentration of motors $K^{\infty}$. Indeed the balance
between unbound and bound motors is given by total particle conservation $N =
N_u + N_b$, with $N_b$ the number of particles attached to the filament.  
When this particle conservation equation is multiplied by $\tilde{\omega}_{\rm
A}/(\omega_{\rm D} V )$, we get:  
\begin{eqnarray}
 K^{\infty} = K + \rho \kappa^\infty,\label{eq:eqcons1}
\end{eqnarray}
with $\rho$  the average fraction of binding sites (on the filament) that are
occupied by motors. We mostly refer to $\rho$ as the bound motor density. The
previous equation can be rewritten as
\begin{eqnarray}
\nonumber
 1 = \frac{K}{K^\infty}  + \rho/m,
\end{eqnarray}
with $m= K^\infty/\kappa^\infty$ as previously defined. When $m\rightarrow
\infty$, $K = K^\infty$  and we recover the TASEP-LK$_{\infty}$ with unlimited
resources as it has been studied in the previous works 
\cite{Par03,parmeggiani2, evans_shock_2003}. The TASEP-LK$_m$ here
introduced is studied in Section 3.

\subsection{Exclusion process in contact with a diffusive reservoir}
\label{sub::mod_diff}
The model presented so far considers one filament in contact with a reservoir
with a homogeneously distributed pool of particles. 
Implicitly, this means that once a binding/unbinding event occurs, particles
diffuse instantaneously in the reservoir to distribute homogeneously, despite
the active transport along the filament that pushes to generate particle
gradients in the solution. In these conditions, the particle attachment rate can
be assumed as constant for any binding site along the filament. 
However, for real systems, particle diffusion is not instantaneous and, in
contrast to the previous scenario, steady-state gradients of particle
concentrations can
build up in the solution. In such a situation, one can no longer assume that the
particle attachment rate is uniform along the filament.  

The second model that we study aims to understand whether diffusion and
concentration gradients have an impact on the
finite resources framework we have introduced with the first model.
As shown in figure 1 (B), we consider that unbound point-like particles diffuse
with a diffusion coefficient $D$ in a two-dimensional box (of area $L_x\times
L_y$). This two-dimensional rectangular reservoir is the simplest generalisation
that considers a spatially extended reservoir of the model previously
introduced. 
While it is possible to consider many different three-dimensional reservoirs and
geometries, for example placing the filament in a cylinder,
the qualitative influence of diffusion on the competition between resources is
not expected to be different in these geometries. 
A quantification of the effects of reservoir geometry on motor transport is out
of the scope of the present work.   Moreover, at
fixed concentrations, adding one dimension needs to be compensated by a
corresponding increase of the number of particles, and therefore a significant
increase of the computational time.

We consider that unbound particles do not interact with each other and their
position $\vec r$ evolves according to 
a Brownian dynamics. The equation of motion is:  
\begin{equation}
\frac{d\vec r}{dt}=\vec\xi(t)\,,
\label{BDEq}
\end{equation}
where $\vec\xi$ is a Gaussian random noise verifying~:
\begin{eqnarray}
 \langle\xi_a(t)\rangle&=&0\,, \label{BDEq2}\\
 \langle\xi_a(t)\xi_b(t')\rangle&=&2D\delta(t-t')\delta_{a,b}\,, \label{BDEq3}
\end{eqnarray}
where the indices $a$ and $b$ stand for the space components ($x$ and $y$
directions). In the simulations, $D=5\mu m^2s^{-1}$, which is the same order of
magnitude as the diffusive constant of motor proteins~\cite{Klumppx,
parmeggiani_non-equilibrium_2009}. The other parameters such as the
particle hopping rate and the size of the reservoir are set to the typical
values of {\it in
vitro} experiments~\cite{Varg2} 
(see table~\ref{table} and \ref{appendix_sim}).
The main difference between this simulation scheme and the one in the papers
\cite{Klumppx,Klumppxx, reviewKlumpp, Muller,
parmeggiani_non-equilibrium_2009},
is that particles in the reservoir diffuse in a continuous space (and not on a
lattice).

An unbound particle can attach to the filament at a constant rate $k$  when it
is located within the reaction volume of a site $i$ of the one-dimensional 
lattice (these are represented by disks in figure 1 (B)).
Therefore, in presence of motor diffusion in the solution, for the reasons
explained above, the attachment rate $\omega_A(i) = \tilde{\omega}_Ac_u(i)$
depends on the position $i$ of the site, through the concentration $c_u(i)$ of
unbound particles present in the reaction volume of the site $i$.
Here  $\tilde{\omega}_A$ is constant for each site, and it is related via
$\tilde{\omega}_A = k v_R$ to the intrinsic attachment rate $k$ of a single
particle and to the reaction volume   $v_R$.  Note that here $v_R$ has the
physical dimension of a 
surface  since the simulations are in two dimensions.  Once attached to the
filament, the particle dynamics on the lattice follows
   the same rules of a standard exclusion process with Langmuir kinetics (as
previously explained in \ref{sub::2.1}). We refer
to \ref{appendix_sim}  for a detailed explanation of the simulation
scheme.

\section{Results: mean field solutions and simulations for TASEP-LK$_{m}$} 
\label{sec::MF}
We now develop an  analytical theory for the stationary density profile of bound
motors and their dependence on  the limited number of motor proteins in the
reservoir (i.e.~resources). The TASEP-LK$_{m}$ process in a finite reservoir, as
defined in section~\ref{sec::22}, has four control parameters: the end
dissociation rate $\beta/p$, the processivity parameter $\Omega_{\rm D}$, the
total concentration of motor proteins $K^\infty$, and the concentration of
tubulin dimers $\kappa^\infty$. 
These control parameters determine two stationary variables: the density profile
of bound motors on the segment $\rho(x)$ and the concentration of unbound motors
$K = c_u\, \tilde{\omega}_{\rm A}/\omega_{\rm D}$. 
The density profile $\rho(x)$ is defined as the continuum limit of the discrete
profile $\rho_i$, with $i=1..L$ and correspondingly $x\in[0,1]$.
Note that the continuum limit corresponds to $L\to\infty$, at fixed value of
$m$, so that the number of particles increases accordingly, i.e.~$N\to\infty$.
The link between the profiles $\rho_i$ and $\rho(x)$ is given by $x=i/L$, and
the stationary current is given by $J(x)=p\rho(x)(1-\rho(x))$. Note  that in
general $\rho_L\neq \lim_{x\rightarrow 1}\rho(x)$ for $L\rightarrow \infty$ due
to the presence of a boundary
layer of finite size (apart from the maximal current phase).  
We will also write $\rho\equiv\int^1_0dx\rho(x)$ for the average density. To
determine the stationary value of $\rho(x)$ and $K$ we consider a set of three
coupled equations determining the essential physics of the model.

The first one expresses the {\it particle conservation}, see equation
(\ref{eq:eqcons1}): 
\begin{eqnarray}
K^\infty= K + \rho \: \kappa^\infty.\label{eq:NConsx}
\end{eqnarray}  
The finite resource parameter defined by $m = K^\infty/\kappa^\infty$ determines
how much $K$ will differ from $K^\infty$, and thus how much the reservoir is
depleted by the presence of the filament. 

The second equation represents the {\it balance of currents between the
reservoir and the filament}: in the stationary state the total current flowing
from the reservoir to the filament is exactly balanced by the current flowing
from the latter to the reservoir.
We get the balance equation:
\begin{eqnarray}
 J_{\rm A} - J_{\rm D} - J_{\rm end} = 0, \label{eq:currentBalance}
\end{eqnarray}
with the attachment current $J_{\rm A}/p = \Omega_{\rm A}(1-\rho)$, the
detachment current $J_{\rm D}/p =
\Omega_{\rm D}\:\rho$ and the end-dissociation current  $J_{\rm end}/p
=[1-\rho(1)]
\rho(1) = 
(\beta/p)\:\rho_{i=L}$ . 
The current balance equation (\ref{eq:currentBalance}) can be solved for
$\rho$,
\begin{eqnarray}
  \rho &=& \frac{K -
(1-\rho(1))\rho(1)/\Omega_{\rm D}}{1+K}.\label{eq:currentConsx}
\end{eqnarray}

The third equation  describes the stationary directed transport process of
motors along the filament. In the continuum limit and in the stationary state,
{\it the equation for the density profile of bound motor proteins $\rho(x)$}
writes generically as \begin{equation}
\partial_x j(x) = \Omega_D\left\{K(1-\rho(x)) -
\rho(x)\right\}\label{eq:densityProfile} \end{equation}
where one can  recognize the local balance between the
current of active particles $j(x)$ and the local binding/unbinding events of
the Langmuir kinetics $\Omega_A(1-\rho(x))$ and  $\Omega_D \rho(x)$. For the
totally asymmetric simple exclusion process the active current writes simply as
$j(x)=\rho(x)(1-\rho(x))$. Note that this equation is actually a mean-field
derivation of the TASEP-LK model where the relevant behavior occurs in the
so-called “mesoscopic limit”, see for a detailed derivation and discussion
 \cite{Par03, parmeggiani2,evans_shock_2003}. 

We consider
two
solutions to the equation (\ref{eq:densityProfile}), a solution $\rho_\alpha$
with the boundary conditions $\rho_\alpha(0) = 0$ and a solution $\rho_\beta$
with boundary condition $\rho_\beta(1) = 1-\beta/p$.  
The analytical expression for the stationary profile
$\rho(x)=\rho\left(x;K, \Omega_{\rm D},\beta/p\right)$ follows by selecting the
solution with the minimal current~\cite{Par03,
parmeggiani2, popkov_localisation_2003}. Here we shortly present the qualitative
results for $\rho(x)$, while the corresponding analytical expressions can be
found in \ref{App:A}. 

As in the TASEP-LK$_{\infty}$, the density profile $\rho(x)$ is characterized by
four different phases: the {\it low-density spike} phase (LDs), the {\it
low-density no jam} phase (LDn), the {\it low density-high density coexistence}
phase (LD-HD) and the {\it low density-maximal current coexistence} phase
(LD-MC). The LDs and LDn phases develop a continuous $\rho(x)$-profile at low
average density
$\rho<1/2$. The distinction between LDs and LDn is defined through the density
value at the last site: in LDs the profile develops a spike at the last site
$\rho_L>\rho(1)$ while in LDn the profile has no spike $\rho_L<\rho(1)$. Note
that the notion of ``spike'' is related to that one of  a ``boundary layer'' in
exclusion processes.  Both spikes and boundary layers characterize a
 significant deviation
of the density profile at the filament end from the bulk profile given by the
continuum limit equations. The main  difference between both notions is that the
spike relates to the average density on the last site while the boundary layer
refers to the density profile over a finite characteristic length from the bulk
towards the last site.

In the LD-HD and LD-MC  phases the profile $\rho(x)$ develops a shock (or
discontinuity) at $x=x_w$. Before the domain wall position (for $x<x_w$) we have
a LD profile, while after domain wall position (for $x>x_w$)  we have a HD or MC
profile. These latter profiles are characterized by a high density $\rho>1/2$. 
The distinction between the HD and MC profile is that the former depends on the
exit rate $\beta$ while the latter is independent of $\beta$ (see appendix A).
The formation of a domain wall in the density profile corresponds to the onset
of a queue of motor proteins on the filament, meaning that the filament is found
in a coexistence of phases, with no jams before the domain wall and the queueing
phase extending from the end of the filament.    
 
To analytically determine the form of the density profile $\rho(x)$ and the
concentration of free motors $K$ at finite resources, we have to solve the set
of coupled equations (\ref{eq:NConsx}), (\ref{eq:currentConsx}) and
(\ref{eq:densityProfile}). In this respect, it is insightful to rewrite $K$ as a
function of the current at the end tip $J_{\rm end} = p(1-\rho(1))\rho(1)$:
\begin{eqnarray}
2K & = & - \left[ 1 + \kappa^\infty (1-m) \right] 
\nonumber \\
\fl &&+ \kappa^\infty \sqrt{\left(\frac{1}{\kappa^\infty} + 1 - m\right)^2
+\frac{4}{\kappa^\infty} \left[m+(1-\rho(1))\rho(1)/\Omega_{\rm D}\right]} \;,
 \label{eq:K1} 
\end{eqnarray}
while $\rho$ as a function of $J_{\rm end}$ is given in equation
(\ref{eq:currentConsx}). Equation (\ref{eq:K1}) gives us readily the reservoir
density in the LD-HD and LD-MC phases where the right boundary is fixed to
$\rho(1) = \rho_\beta(1) = (1-\beta/p)$ and therefore $\rho(1)$ is independent
from $\rho$
and $K$. We have then that $\rho(1) = 1-\beta^\ast = 1-{\rm
min}\left\{\beta/p, 1/2\right\}$ and we get the explicit expression for the
rescaled reservoir density $K$ in the LD-HD and LD-MC phases:
\begin{eqnarray}
2K & = & - \left[ 1 + \kappa^\infty (1-m) \right] 
\nonumber \\
\fl &&+ \kappa^\infty \sqrt{\left(\frac{1}{\kappa^\infty} + 1 - m\right)^2
+\frac{4}{\kappa^\infty} \left[m+(1-\beta^\ast)\beta^\ast/\Omega_{\rm
D}\right]}.
\label{eq:K1x}
\end{eqnarray} 
When the filament is in the LD phase we do not have such an explicit
expression  for $\rho(1)$ that is no longer independent of $K$, and we are lead
to
solve an implicit equation on $K$.  As discussed in detail in~\ref{App:A}, it is
possible to write
an
expression for $\rho(1)$ as a function of $K$;  plugging this expression into
equation~(\ref{eq:K1}) then provides for the implicit solution equation.  This
approach can be developed also in presence of a finite homogeneous
reservoir with excluded volume interactions between particles, see
section~\ref{subsec:depletion}. \\

We now present the results of the mean field framework presented in this
subsection: first we
discuss the depletion effects of the reservoir as a function of the finite
resource parameter $m$, and then we show how finite resources affect the
formation of jams; finally we present the novel phase diagrams obtained for
various values of $m$.  

\subsection{Depletion of the reservoir}\label{subsec:depletion}
\begin{figure}[b]
\begin{center}
\includegraphics[ width =1\textwidth]{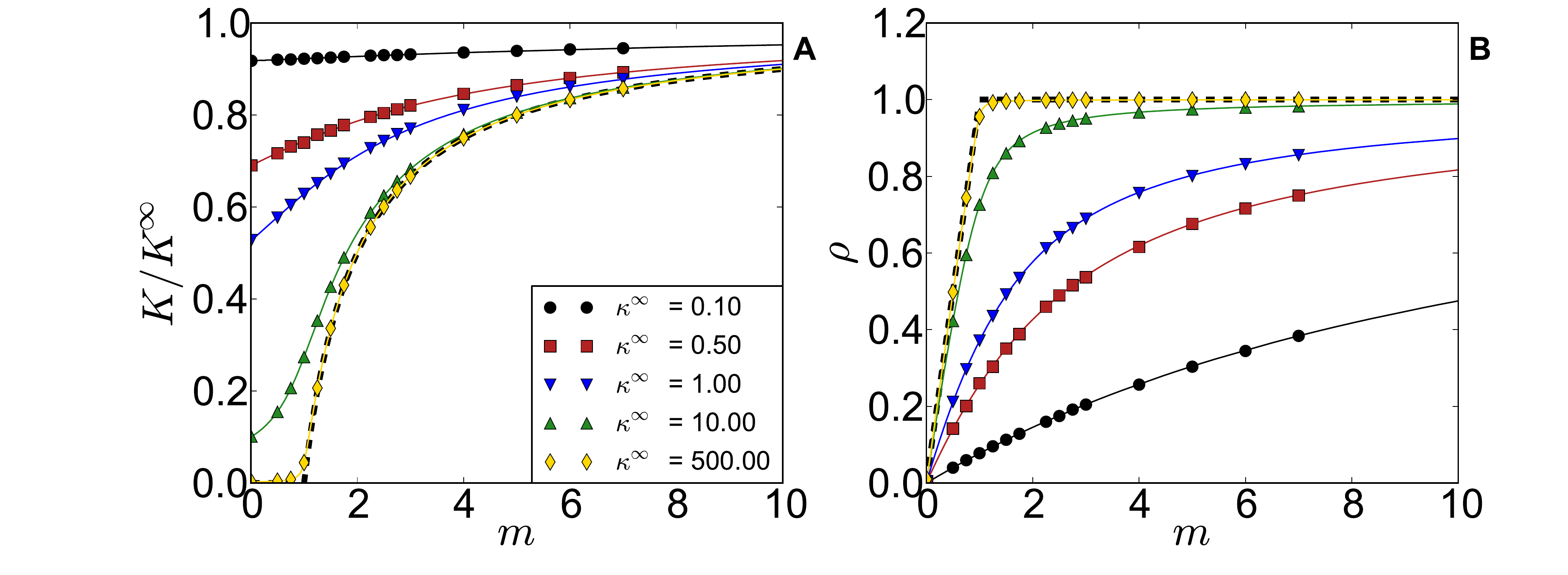}
\caption{The depletion of particles in the reservoir $K/K^\infty$ (A) and the
filling 
 of the segments $\rho$ (B) as a function of $m$ for given values of 
$\kappa^\infty$, with $\Omega_{\rm D}=10$ and $\beta/p = 0.8$. Solid lines are
analytic expressions from mean field, Eqs.~(\ref{eq:currentConsx}) and
(\ref{eq:K1}), and symbols are obtained from numerical simulations
(see~\ref{appendix_sim} for details). The black dashed line (which is almost
superimposed by the yellow line with diamond markers)  represents the limit
$\kappa^\infty
\rightarrow \infty$ of Eqs.(\ref{lim:kappa_inf}) and
(\ref{lim:kappa_inf2}).}
\label{fig:finite1_c}
\end{center}
\end{figure}

In figure~\ref{fig:finite1_c} we have plotted the ratio $K/K^\infty$, between
the motor
concentration in the reservoir with respect to their total
concentration, and the density of bound particles $\rho$ as a
function of the finite resource parameter $m$. In particular, these curves can
be interpreted
as varying the number of motors $K^{\infty}$ while keeping the concentration
$\kappa^\infty$ of tubulin dimers fixed, or reversely varying the concentration
of tubulin dimers while keeping the concentration of motors fixed.
We observe how the depletion of motors from the reservoir becomes relevant for
small $m$ and large tubulin dimer concentrations (figure~\ref{fig:finite1_c}
(A)).
Indeed, for an infinitely large reservoir we should have that $K^\infty=K$ while
for small values of $m$ we get a $K\ll K^\infty$ indicating the depletion of the
reservoir. Depleting the reservoir leads as well to a smaller density of bound
motors since the stationary attachment rate decreases. 

At constant $m$, in the limit of  abundant binding sites and high motor
concentrations we can analytically 
characterize the
depletion effects.  In the asymptotic limit, at which the motor protein
concentration $K^\infty$ and the concentration of binding sites $\kappa^\infty$
are infinitely large, we find
\begin{eqnarray}
\lim_{\kappa^{\infty}\rightarrow \infty} \frac{K}{K^\infty} =
\left\{\begin{array}{ccc}0 && m\leq m_c = 1 \,,\\ 
 (m-1)/m &&  m > m_c = 1\,,
        \end{array}\right. \label{lim:kappa_inf} \\ 
\lim_{\kappa^{\infty}\rightarrow \infty} \rho = \left\{\begin{array}{ccc} m &&
m\leq m_c = 1 \,,\\ 
                                                                                
                                        1  && m > m_c = 1 \,.
\end{array}\right. \label{lim:kappa_inf2}
\end{eqnarray} 
Interestingly, we recover a continuous phase transition in the reservoir
consumption
$K/K^\infty$. At very high levels of tubulin dimer concentrations, the reservoir
is strongly depleted for values of $m$ smaller than $m_c = 1$. By increasing
$m<m_c$, most of the particles introduced into the system will add to the
density of bound particles $\rho$, which thus increases linearly with $m$ until
it eventually saturates and the filament is almost completely filled (see
the yellow line with diamond markers in figure~\ref{fig:finite1_c} (B)).

Equation (\ref{lim:kappa_inf}) implies that $K$ will have a finite value for
$m<m_c$, while it becomes infinite large for $m\geq m_c$.  The transition at
$m=m_c$ can by characterized by a change in the
algebraic behaviour of the motor protein concentration $K$ in the reservoir as
a function of the  concentration of binding sites $\kappa_\infty$, viz.~:
\begin{eqnarray}
 \fl K =
\left\{\begin{array}{ccc}\frac{m}{1-m}\left(1+
\frac{\rho(1)(1-\rho(1))}{m\Omega_D}\right)  +
\mathcal{O}\left[\left(\kappa^\infty\right)^{-1}\right]  && m< m_c \,,\\  
m\sqrt{\kappa_\infty}\left(\sqrt{1+(1-\rho(1))\rho(1)/\Omega_{\rm
D}}\right) + \mathcal{O}\left[1\right] && m=m_c =
1\,, \\ 
 (m-1)\kappa_\infty  +
\mathcal{O}\left[1\right] && m >
m_c\,.\end{array}\right. \label{lim:kappa_infx} 
\end{eqnarray} 
We thus have for high values of $\kappa_\infty$ that the motor protein
concentration in the reservoir converges to the finite value $m/(1-m)\left(1+
\frac{\rho(1)(1-\rho(1))}{m\Omega_D}\right)$ for $m<m_c$, while it
increases proportionally to the binding site concentration $\kappa_\infty$ for
$m>m_c$.  For values $m=m_c$ the motor protein concentration increases as a
square root of  $\kappa_\infty$. 

As it will become clear in the next subsection, by measurements of the
concentration of motors $K$, it is possible to determine the stationary density
and current of the filament. Moreover, here we suggest that the motor
protein density
 on a  filament and its dependence on the reservoir concentration can depend on
the local level of filament crowding
(i.e.  with different $\kappa^\infty$).  This suggests that distinct parts  
of the cytoskeleton with locally different levels of
filament   crowding  can potentially show rather different transport
characteristics.
This  aspect  is rationalized by the dependence of the derivative of the 
density, $\partial_m \rho$, see figure \ref{fig:finite1_c} (B). This
derivative can strongly depend  on $\kappa^\infty$, i.e. on the local 
concentration of potentially available binding sites. Interestingly, this
argument can apply also to networks of filaments in
contact with the finite  reservoir of particles.

\subsection{Density profiles by varying supply and
demand}\label{subsec:densprofile}
\begin{figure}[t]
\begin{center}
\includegraphics[width=0.99\textwidth]{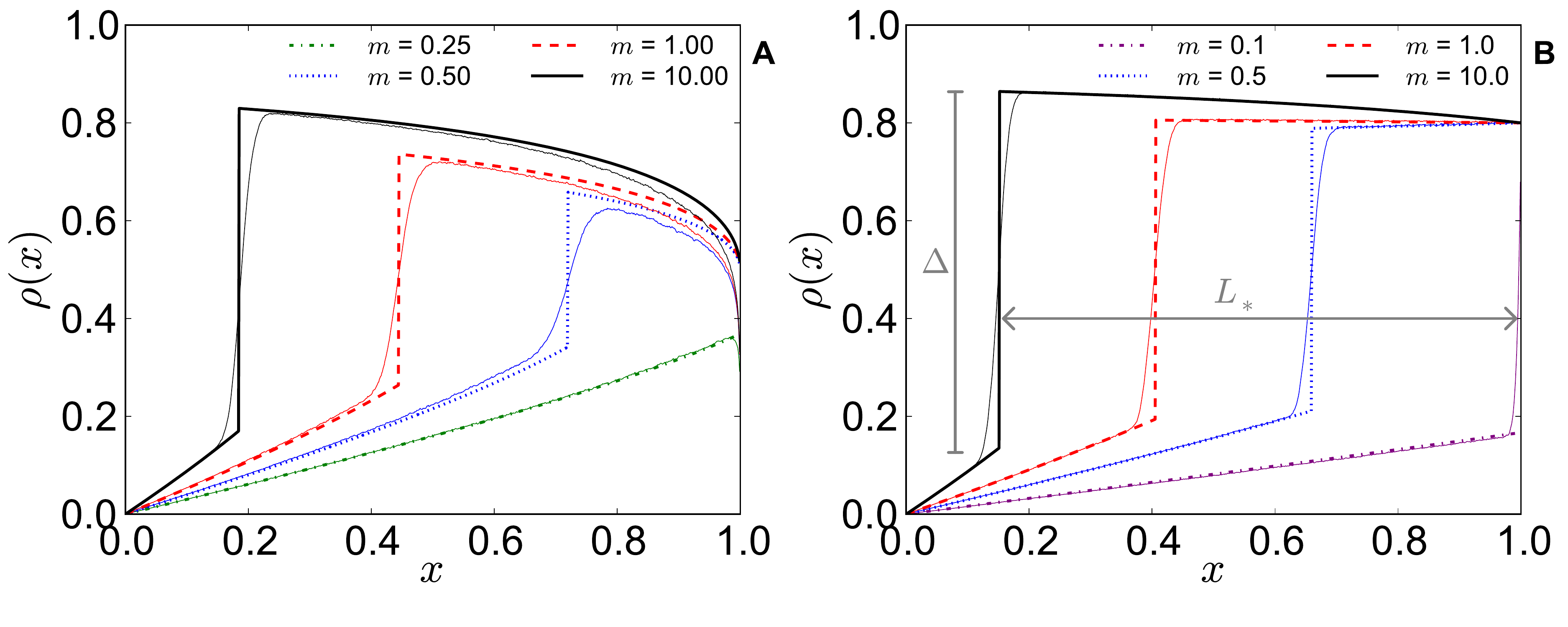}
\caption{Density profiles for a TASEP-LK$_{m}$ for different values of
 $m$ (see legend). In panel (A) we present a system with $\beta/p=0.8$,
$\rho_l^\infty=0.9$ and $\Omega_{\rm D}=0.1$, moving from a LD-MC phase to a LDn
phase when the resources become limiting (small $m$). Panel (B) shows the same
system with $\beta/p=0.2$, with a LD-HD phase for large $m$ and a LDs for small
$m$. Light lines are the outcome of numerical simulations with $L=1000$. We
observe (not shown here) that the width of the domain wall becomes
thinner as
$L$ increases in the simulation results. We  also indicate in the figure the
domain wall height $\Delta$
and the domain wall distance from the filament end $L_*$.}\label{fig:finite4}
\end{center}
\end{figure}
Recall that the finite resource parameter $m$ indicates the number of
motors in solution, relative to the number of polymerised tubulin dimers.
In figure~\ref{fig:finite4} we present the variation of the density profile
$\rho(x)$ for given
values of $\Omega_{\rm D}$, $\beta/p$ and total concentration of motors
$K^\infty$ as a function of $m$. 
Decreasing $m$ reduces the observed density of processing motors on the
filament, because less motors are available in solution.   Thus the number of
reservoir motors per binding site decreases.

\begin{figure}[h]
\begin{center}
\includegraphics[width =0.99\textwidth]{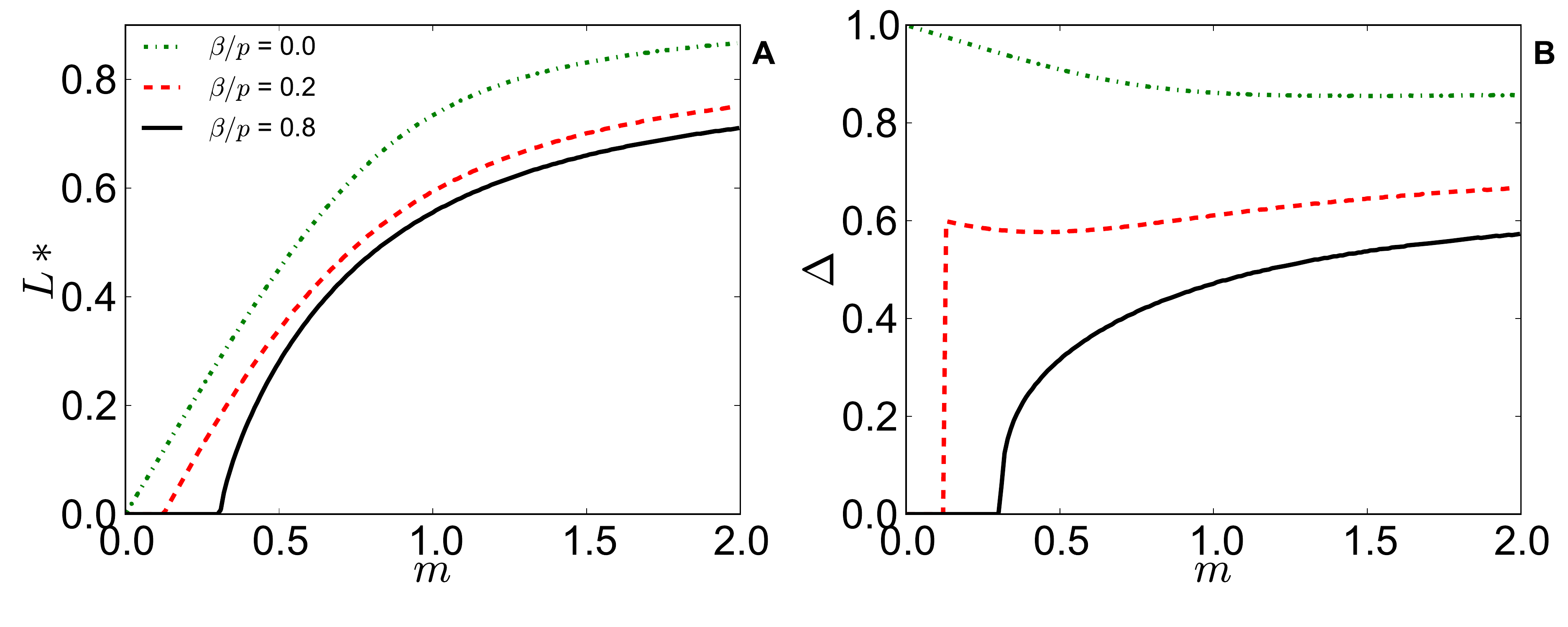}
\end{center}
\caption{The jam length $L_\ast$ and domain wall height $\Delta$
are presented as a function of $m$ for $\rho^\infty_\ell=0.9$, $\Omega_D=0.1$
and
given values of $\beta/p$. The curves are obtained from the mean field theory;
black full lines denote a LD-MC to LDn transition, while the other curves show a
LD-HD to LDs transition.
}\label{fig:jamlength}
\end{figure}

With our approach we can also
study the features of the motor protein jams. Figure \ref{fig:finite4} shows
that a reduction of the resource parameter $m$ shifts the position of the domain
wall to the right and thus decreases the jam length $L^\ast$ (the jam length is
defined as $L_\ast=1-x_w$, with $x_w$ the domain wall position). 
This is an experimentally accessible quantity~\cite{Varg2} that might be
exploited to compare models and experimental outcomes.
In figure \ref{fig:jamlength} (A) we characterize the decrease of the jam length
as
a function of $m$. The jam length $L^\ast$
 reaches zero length at a critical value of $m$ for which the system undergoes a
phase transition from a LD-HD phase (or LD-MC) to a LDs (or LDn) phase. 
In figure~\ref{fig:jamlength} (B) we present the domain wall height $\Delta
=1-2\rho(x_w)$ as a function of $m$. Contrary to the the jam length
$L_\ast$, the domain wall height $\Delta$ does not necessarily decrease as a
function of
$m$.  
Instead,  we observe a different behaviour depending on whether the
system is initially found in the LD-HD or LD-MC phase. In fact, the shock in the
LD-MC disappears with a continuous transition in $\Delta$, while the shock in
LD-HD disappears with a discontinuous transition in $\Delta$. 
So, whether the profile is in LD-HD or in LD-MC will determine its qualitative
behaviour as a function of $m$.

Theory is  compared to numerical simulations in figure~\ref{fig:finite4}; our
analytical results are in good agreement with the simulations
and the shock
profile from the numerical simulations shows the expected finite-size effects. 
The slight disagreement between simulations and mean field solutions in the MC
region is also present in the TASEP-LK$_{\infty}$~\cite{Par03, parmeggiani2} and
in  TASEP~\cite{krug_boundary_1991,schutz_phase_1993}, and it is due to the
divergence of the length scale of the boundary layer in the MC phase.

\subsection{Phase diagrams at finite resources}
In the previous subsection we have shown that by varying the parameter $m =
K^\infty/\kappa^\infty$ we can induce phase transitions in the motor density
profile
on the filament. In this subsection we  complement the picture by
presenting how the phase diagram of TASEP-LK$_{m}$ depends on the finite
resource parameter $m$.  
Substituting  the solution $K$ to the equations
(\ref{eq:currentConsx}), (\ref{eq:densityProfile}) and (\ref{eq:K1}) into the
equations for the phase transition lines of TASEP-LK$_{\infty}$ presented 
in \ref{app:infRes} determines such a phase diagram.  Further detailed
analysis is presented in
\ref{sec:phaseTransFinite}. 
Here we  focus on the main results.  

\begin{figure}[t]
\begin{center}
\includegraphics[width =0.99\textwidth]{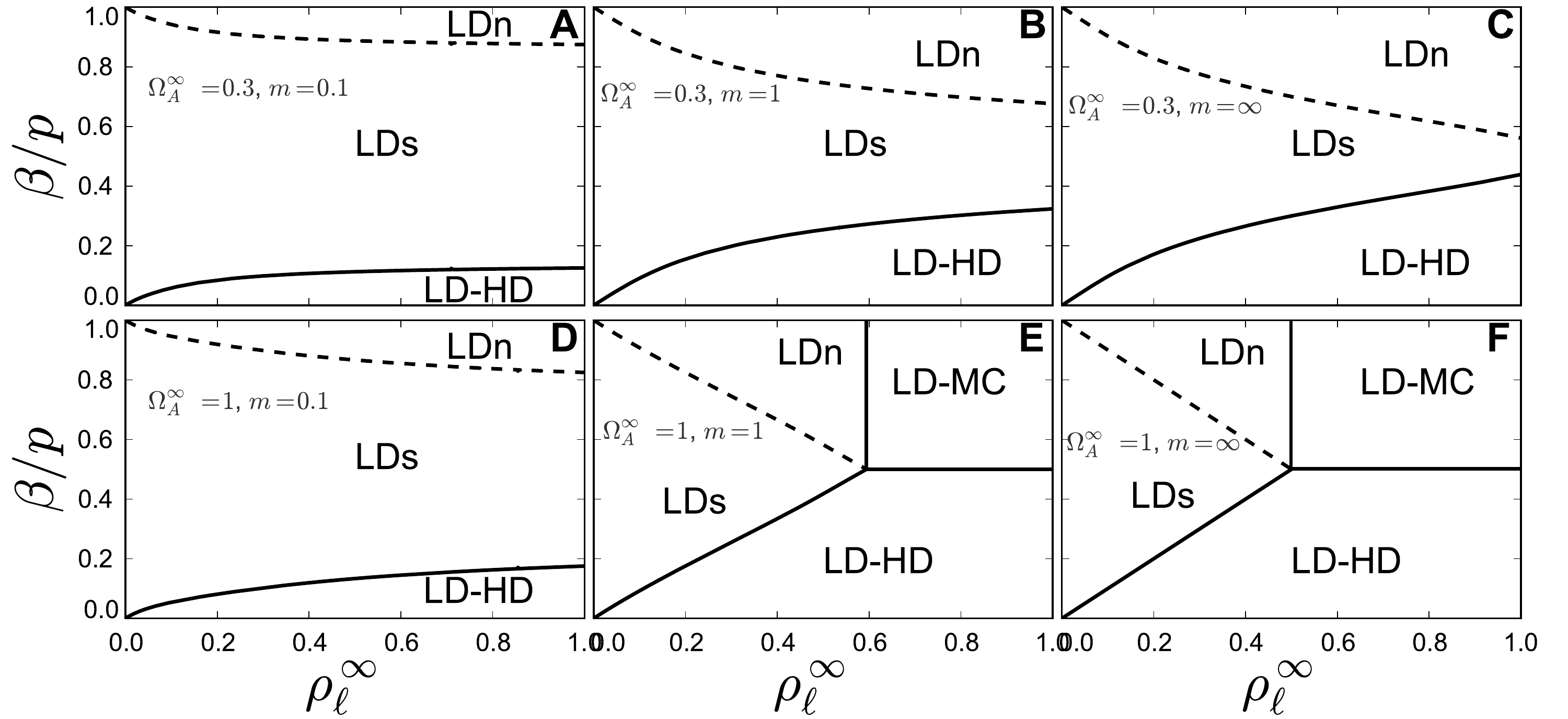}
\caption{The $(\rho_\ell^\infty, \beta/p)$-phase diagrams of TASEP-LK$_{m}$ at
given values of $\Omega^\infty_A$ and $m$.  When $m$ becomes large we recover
the phase diagrams presented in~\cite{Varg2}.
 }\label{fig:finite2}
\end{center}
\end{figure}

In figure \ref{fig:finite2} we present the $(\rho^{\infty}_\ell,\beta/p)$ phase
diagram
for certain fixed conditions $(m, \Omega^\infty_A)$. This type of
phase diagram in the parameters $(\rho^{\infty}_\ell,\beta/p)$ has been
introduced in~\cite{Varg2}.   Upon decreasing   $m$ the LDs phase
gradually
dominates the  diagram and the LD-HD and LD-MC phase regions reduce in size. The
LD-HD phase disappears only for $m\rightarrow0$, whereas there exists a critical
$m$ such that the LD-MC is no longer attainable (see
\ref{sec:phaseTransFinite}). Hence, when the motor resources are
limited, the stationary density profiles are dominated by spikes. 
  This is compatible with the behaviour shown in figure \ref{fig:finite4} (B):
by
decreasing $m$,  the density profile exhibits signatures of a LDs. This result
is
also consistent
with the limit $m \rightarrow 0$, corresponding to non-interacting isolated
motors moving along the filament. Then the density profile is proportional to
the residence time of the motors on the lattice dimers. In the $m\rightarrow 0$
limit, with
filaments  sparsely populated by motor proteins, our model reduces to
the ``antenna model'' that has been considered in~\cite{VargaCell}. Due to
the spike being located on the last site, this feature of the LDs might be
important to regulate filament depolymerization by motor
proteins~\cite{VargaCell, klein_filament_2005, Johann, Melbinger, Kuan}. In
fact, the length-regulation of microtubule filaments is determined by the
transit of motor proteins through their end tip and hence to the density at the
end of the filament.  
It would therefore be interesting to study how the changes in the TASEP-LK phase
diagram  affects the depolymerization rate at finite resources, and hence if
competition for resources can also determine the characteristic filament
length. 

The diagrams in figure~\ref{fig:finite2} also allow us to compare the present
study
with the results presented in the experimental work~\cite{Varg2}, where
measured density profiles are interpreted using the revisited phase diagrams
of TASEP-LK$_{\infty}$. In particular, we have chosen 
$\Omega^\infty_{\rm A}=0.3$, as estimated in~\cite{Varg2}, to facilitate the
comparison of our
diagrams with the figures presented in~\cite{Varg2}. 
We stress that, according to our estimates, $m\approx 10^3$ in \cite{Varg2} ,
meaning that the system is found in the infinite reservoir limit.  If $m$ were
reduced, e.g.~by reducing the motor protein concentration while keeping the
administered tubulin constant, we predict that the LDs phase would dominate the
resulting experimental profiles.  To check this, we suggest to repeat the assay
\cite{Varg2}, moreover reducing the employed motor protein concentration to
approximately $1pM$.  

\begin{figure}[h]
\begin{center}
\includegraphics[width =0.99\textwidth]{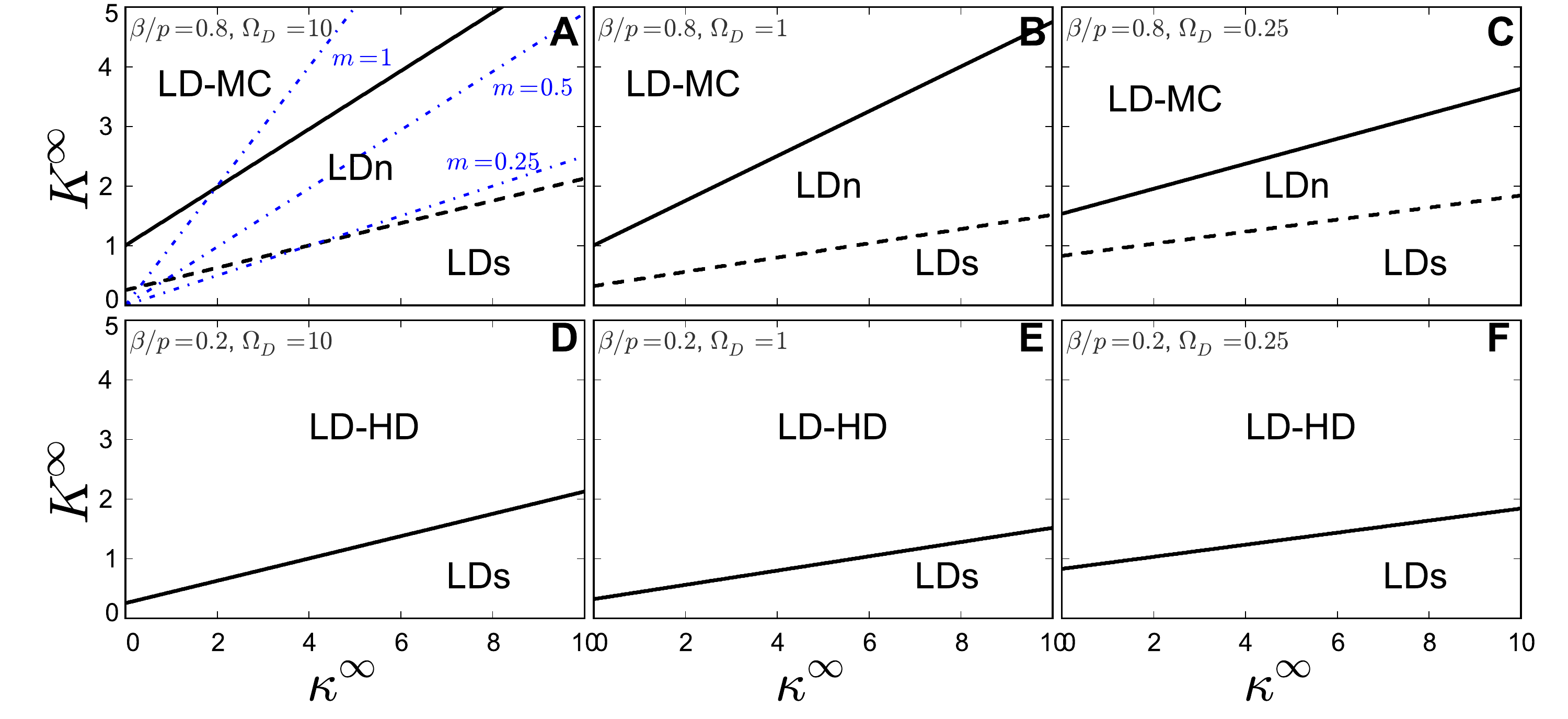}
\caption{$(\kappa^\infty, K^\infty)$-phase diagrams for TASEP-LK$_{m}$ at given
values of $\Omega_{\rm D}$ and $\beta/p$. 
Thin dashed curves in panel (A) represent lines of constant $m$ ($m=0.25$,
$0.5$, $1$).}
\label{fig:finite3}
\end{center}
\end{figure}

An alternative way to represent the phase diagrams allows one to explicitly
follow the development of phases as a function of the motor
protein and tubulin concentrations in solution. In figure~\ref{fig:finite3} we
present such $(\kappa^\infty,K^\infty)$-phase diagrams at fixed $\Omega_{\rm D}$
and $\beta$, where the motor concentration $K^\infty$ and the concentration
of polymerized tubulin dimers $\kappa^\infty$ are the employed control
parameters.  The latter gives rise to the possibility to fully sample the phase
diagram \cite{Varg2} with the practical advantage that neither motor
processivity nor end dissociation need to be controlled, e.g.~by adjusting the
solution's salt concentration.  

Tracing  dot-dash  lines in these $(\kappa^\infty,K^\infty)$
diagrams
exhibits the system's phase
behaviour for a fixed value of $m=K^\infty/\kappa^\infty$. Importantly,  that
there is no 
direct way to pass from a LD-MC  to an LDs phase. Thus, by
decreasing the motor concentration  $K^\infty$ the system will necessarily pass
through the LDn phase.
Starting from the LD-HD regime, the system can only undergo a transition to a
 LDs phase, see figure \ref{fig:finite3} panels (D-F). 
We also emphasize that the location of the transition lines depend only weakly 
on the
processivity $\Omega_D$: between the first and last column of
figure~\ref{fig:finite3}
the processivity varies by a factor $40$, without affecting the qualitative
location of
the transition lines.   

These phase diagrams might hence be  relevant in experimental scenarios
like~\cite{Varg2}. Since the total concentration of
motors $K^\infty$ or of bound tubulin dimers $\kappa^\infty$  are
  controllable, these
phase diagrams could allow for a quantitative interpretation of
depletion effects on the observed density profiles.  
A typical feature appearing only at small $m$ is the
predicted large extension of the LDs phase with respect to all phases.  If the
suggested experiments would manage to detect spikes in the motor density, one
could hope for an accurate quantification of the prevalence of LDs over  LD-HD
and LDn.

\subsection{Filaments without end dissociation rate} 
We end the analysis of the first model with the particular case for $\beta=0$
that allows us to derive some
general model-independent results. In this case the current leaving the end of
the tip is zero, and equations (\ref{eq:NConsx}) and (\ref{eq:currentConsx})
decouple.  The profile is in the LD-HD phase and the density at
the end of the filament $\rho(1) = 1$, is independent of the concentration of
unbound motors $K$~(see
\ref{app:infRes}). We thus have $\rho = \rho_\ell \equiv K/(1+K)$, with a $K$
that follows readily 
by setting $\rho(1)(1-\rho(1))=0$ in equation (\ref{eq:K1}). 
In the $\beta=0$ case, we recover the
results presented by Klumpp and Lipowsky in~\cite{Klumppxx}. This is not a
coincidence, as both results are derived using the same two arguments: the total
conservation of particles in the system and the balance of the currents between
reservoir and filament.  
The results can be linked using a general formula we provide in~\ref{App:B}. 
The average density $\rho$ of bound motors for $\beta=0$ applies to a much
larger class of transport processes
than TASEP-LK$_{m}$ (although the shape of the profile  $\rho(x)$
is model dependent).     Indeed, this is a direct consequence of the fact that 
when $\beta=0$ we do not need to consider the
equation (\ref{eq:densityProfile}) for the steady-state density profile on the
filament to solve equations (\ref{eq:NConsx}-\ref{eq:currentConsx}) and the
density on the filament is determined solely by the Langmuir adsorption process
\cite{Langmuir18}. 
Consequently, our results remain valid for any other type of microscopic dynamic
rule for the motor protein motion on the filament which does not correlate
with the detachment/attachment process (such as bi-directional
motion, for example) and they  hold even for configurations
with multiple filaments immersed in a homogeneous reservoir.  
For instance, the formalism proposed in this section is also valid for a network
of cross-linked filaments or branched filaments. Moreover, the stationary
density profile for a TASEP-LK$_{m}$ on a network can be determined using the
results~\cite{NeriTT, NeriTTT} and considering the appropriate value of $K$ (and
thus $\rho$), as given by equations (\ref{eq:currentConsx})-(\ref{eq:K1}) when
setting $\rho(1)=1$.  
\footnote{In this case the condition $\rho(1)=1$ means that
the current exiting the end of each filament is 0, or in other words that the
network is ``closed'' (no exit from the end tips).}

\section{Simulation results for TASEP-LK in a diffusive reservoir}
\label{sec::diff}
When not attached to a filament, motor proteins in a flow chamber or in the
cytosol diffuse in solution with a presumed diffusion
constant $D$.
In stationary conditions, a gradient of motors is
necessary to create the
 diffusive current  balancing the current
on the filament.
By considering a diffusive reservoir, the motor concentration of particles
$c_u(i)$ becomes a function of the site position $i$. 
\begin{figure}[h]
\begin{center}
\includegraphics[width =0.99\textwidth]{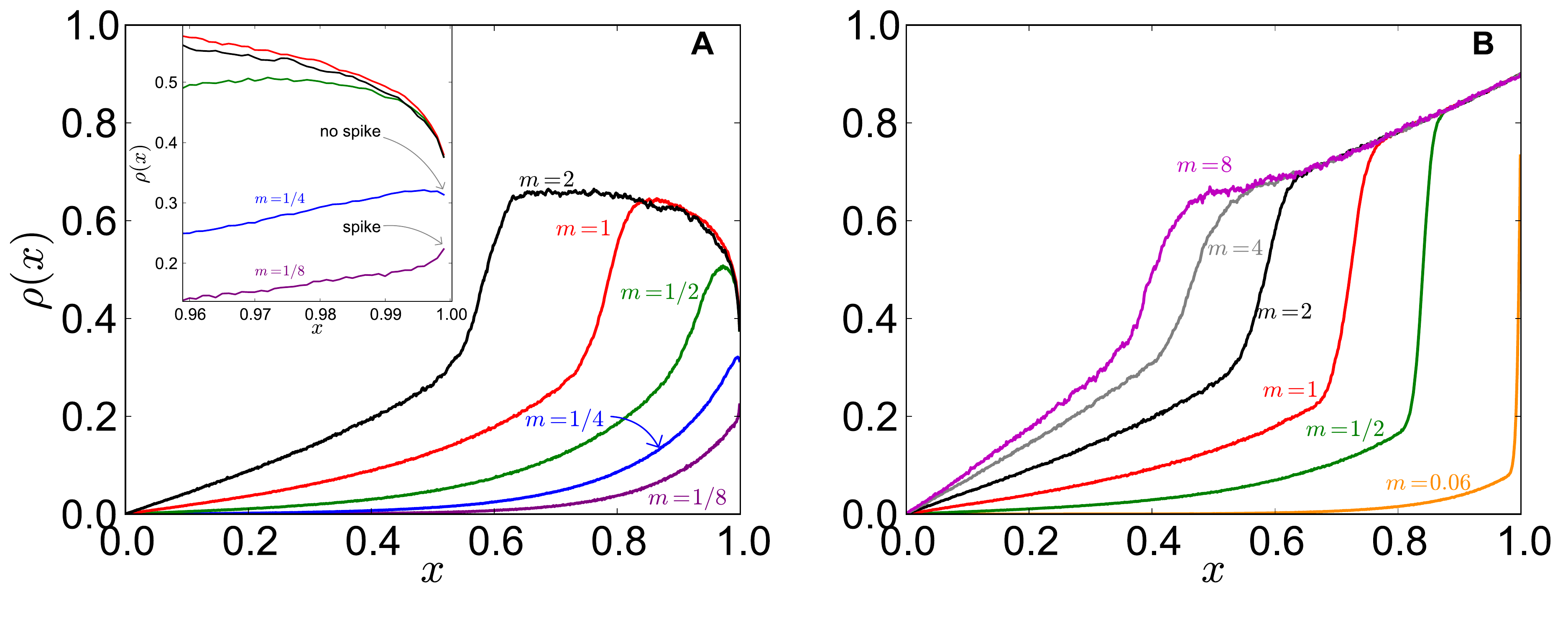}
 \caption{Density profile $\rho(x)$ obtained by explicitly considering the
diffusion of particles in the reservoir. We have considered different values of
$m$ starting from LD-MC (panel (A), $\rho_\ell^\infty=0.7$, $\beta/p=0.7$) and
LD-HD (panel (B), $\rho_\ell^\infty=0.6$, $\beta/p=0.2$). In both panels we take
$\Omega^\infty_{\rm A}=1$, $L=10^3$, the filament hopping rate $p=50s^{-1}$, the
diffusion coefficient $D=5\mu
m^2s^{-1}$, and the reaction volumes are modelled by spheres with diameter
$a=10^{-2}\mu m$.  The dimensions of the reservoir
 are $L_y =0.5\mu m$ and $L_x =100\mu m$.  
We observe the same phenomena as in TASEP-LK$_m$: for large $m$ there appears a
domain wall in the system.  In panel (A) there is a transition from LD-MC at 
$m =$1, 1/2 over LDn at $m=1/4$ to LDs at $m=1/8$.  In panel (B), the system
goes from LD-HD at $m=2$ to LDs for $m\lesssim0.06$. 
\label{fig::diff0}}
\end{center}
\end{figure}
Here we show that a TASEP-LK with a diffusive reservoir exhibits the same
phenomenology as a TASEP-LK$_m$ without reservoir motor diffusion. 

To this end,  we perform numerical simulations of the model introduced in
Section~\ref{sub::mod_diff}, providing a simplified and {\it in silico} version
of the experiment presented in~\cite{Varg2}. The details and the parameters of
the
employed simulation scheme can be found in~\ref{appendix_sim} and in
table~\ref{table},
respectively.
\begin{table}
\caption{\label{table}Parameters of the model with ``diffusive`` reservoir:
values estimated from \textit{in
vitro} experiments~\cite{VargaCell,Varg2} (if not otherwise specified) and
values used in simulations throughout the paper. The last row shows the
competition parameter $m$, which has been fixed to high values in \textit{in
vitro} experiments of kinesin transport along microtubules~\cite{Varg2} to avoid
depletion effects; here we can modulate the parameter $m$ to study the
competition for resources.}
\begin{indented}
\lineup
\item[]\begin{tabular}{@{}l p{4.7cm} l p{3.3cm}}
\br                              
&& Typical values in \\&& experimental conditions & Simulations \\
\mr
$D$ & Diffusion coefficient &$\approx4~\mu m^2s^{-1}$~\cite{Klumppx,
blasius_recycling_2013} &$5~\mu m^2s^{-1}$\\
$p$  &Hopping rate &$1-100~s^{-1}$~\cite{schnitzer_kinesin_1997} & $50~s^{-1}$\\
$\beta/p$ & Rescaled end-dissociation rate & $0.01 - 0.95$ & $0.1 - 0.9$\\
$a$  & site length & $8~nm$ &$10~nm$\\
\multirow{2}{*}{$L_{x,y}$}&\multirow{2}{*}{Reservoir edge-sizes} &
\multirow{2}{*}{$\approx 10^2\mu m$} & $L_x=12.5-200~\mu m$; $L_y=0.25-4~\mu
m$\\
$L$ & Microtubule length & $\sim$250-1250 sites & 1000 sites\\
$\Omega_A^\infty$ & Rescaled association rate$^{\rm a}$ & $\sim0.3-1$ &
$0.3-1$\\
$\rho_\ell^\infty$ & Langmuir density$^{\rm a}$ &$\sim0.1-0.9$ & $0.25 - 0.7$\\ 
$m$ & Competition parameter &$\gtrsim 1000$$~^{\rm b}$ & $\sim0 - 10$\\ 
\br
\end{tabular}
\item[] $^{\rm a}$ To better emphasize that the range of values used in
simulations is of the same order of magnitude of the experimental ones, in this
table we have used $\Omega_A^\infty = c\,\tilde{\omega}_A L/p = K^\infty
\Omega_D$ and $\rho_\ell^\infty = K^\infty/(K^\infty+1) =
\Omega_A^\infty/(\Omega_A^\infty +  \Omega_D)$ defined in Section~\ref{sub::2.1}
as control parameters of the model.
\item[] $^{\rm b}$ Value estimated from~\cite{VargaCell, Varg2}. 
\end{indented}
\end{table}

In figure~\ref{fig::diff0}-(A) we show an LD-MC density profile whereas  in
figure~\ref{fig::diff0}-(B) we show an LD-HD 
density profile that are obtained by model simulations for various ratios $m$.
As observed in TASEP-LK$_m$, the position of the domain wall
is strongly affected by changing the competition parameter $m$.   
Decreasing the value of $m$ leads to a gradual shift in the domain wall towards 
the end of the lattice.
In panel (A), starting from an LD-MC profile and decreasing $m$, the system
undergoes first a transition to a  LDn profile without spike and eventually 
to a LDs profile with a
spike, see the inset of figure~\ref{fig::diff0}-(A). This is an interesting and
prominent feature predicted by the
TASEP-LK$_m$ phase
diagram  from figure~\ref{fig:finite3}.  Since
experiments can detect the appearance of spikes, the domain wall dependence on
controllable parameters allows for an experimental validation of the models
studied. 
In panel (B), starting from the LD-HD phase and decreasing
$m$, the
domain wall disappears and the profiles display the signature of  LDs with a
spike at the end of the filament, see $m=0.06$ in figure~\ref{fig::diff0} (B).
Moreover, in the panels of figure \ref{fig::diff0} we observe a dependency of
the length $L_*$ and the
height $\Delta$ of the shock on the supply-demand parameter $m$ that is in
agreement with the one noticed in the TASEP-LK$_m$, see
figures~\ref{fig:finite4} and \ref{fig:jamlength}.  By decreasing $m$, the
length
$L_*$ of the jam decreases and the height $\Delta$ goes abruptly to zero
(leading to a spike)
or smoothly to zero when starting from the LD-HD or LD-MC respectively.

These results show the robustness of the observations made for
TASEP-LK$_m$:  the presence of diffusion and
 motor concentration gradients does not affect the qualitative features of the 
transitions
from LD-MC to LDn and from LD-HD to LDs.   

\begin{figure}[b]
\includegraphics[width =0.99\textwidth]{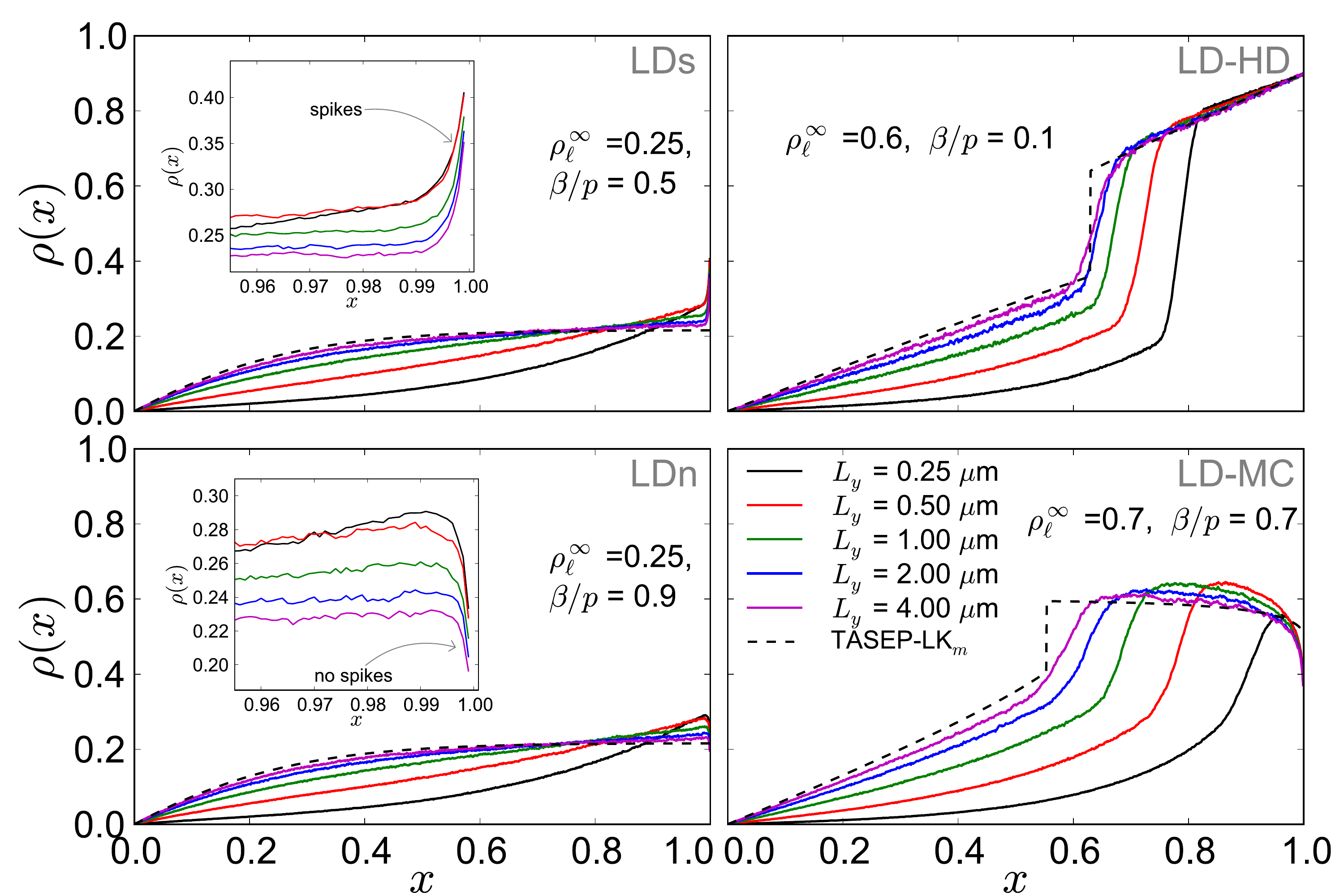}
 \caption{Density profile of motors along the filament for the TASEP-LK in a
reservoir with diffusive particles.     The different panels correspond to the
phases explained in figure~\ref{fig:finite2}, and in particular panel (E).
When the section $L_y$ of the reservoir is increased (from 0.25 to 4.00 $\mu$m)
we observe the convergence of the profiles
to the mean field value expected for the TASEP-LK$_m$ (dashed line). This
corresponds  to a concentration gradient along the longitudinal direction of
the filament that becomes negligible (not shown).  We kept hereby the volume
 $L_x\cdot L_y=50\mu m^2$ fixed and
left all the other parameters unchanged.  
 In the insets of the left
panel we magnify the region close to the end of the filament to emphasize the
presence of spikes (LDs) or their absence (LDn). All the figures have
$\Omega^\infty_{\rm A}=1$, $m=1$,
 $L=10^3$, the hopping rate on the filament $p=50s^{-1}$, the diameter of the
site reaction volume is $a=10^{-2}\mu m$, $c=3.32\cdot10^{-11}mol\,m^{-2}$ and
$D=5\mu m^2s^{-1}$.}
\label{finite_diff1}
\end{figure}

In figure~\ref{finite_diff1} we show that, remarkably, the exclusion model with
diffusion converges to the analytical solutions of the TASEP-LK$_m$
under certain conditions.
In particular, we have checked the effects of  reservoir size on the density
profile on the filament. As we show in figure~\ref{finite_diff1}, when the
cross-section $L_y$ of the reservoir increases, the density profile converges
toward the mean field prediction of the TASEP-LK$_m$. This has an intuitive
explanation: in a closed compartment, the diffusive motor current passing
through a
surface perpendicular to the filament counterbalances the opposite active
current along the filament.  As explained below, the
diffusive current is proportional to both the cross-section
and the gradient of particles in the reservoir.  Accordingly, increasing $L_y$
will decrease
the concentration gradient.  

More specifically, the current of motors $J_f$ on the
filament is in general not homogeneous for the TASEP-LK, and depends on the
position $x$ i.e.~$J_f=J_f(x)$. The opposite current $J_r$ in the reservoir is
related to
the concentration gradient $\vec{\nabla} c(x,y)$, where the concentration is now
a function of the position in the plane $(x,y)$.    We  have checked numerically
for the largest edge-size in the $y$ direction  ($L_y\sim4\mu m$) that the
concentration of particles is  homogeneous along the $y$ direction. 
Homogeneity for the parameters considered is thus expected to  hold as well at
smaller edge-sizes $L_y$. 
Therefore we can assume that the concentration is homogeneous along the
y-direction i.e. $c(x; y) = c(x)$. The same  reasoning would hold for a third
spatial dimension $z$.

The diffusive current in the reservoir reads $J_r(x)=-DL_y\,dc(x)/dx$.
Balancing the
currents  $J_f(x)=J_r(x)$  at position $x$ implies:
\begin{eqnarray*}
J_f(x)&=&DL_y\frac{dc(x)}{dx}\,.\label{balJ}
\end{eqnarray*}
This equation tells us that, for a given (upper bounded) current on the
filament, the gradient vanishes
upon an increase of the cross-section $L_y$. For the largest cross section $L_y
= 4\mu m$,
the motor gradients is typically $16$ times smaller as compared to the
smallest cross section $L_y = 0.25\mu m$. Hence the finite diffusion in the
reservoir
is counterbalanced by appropriate dimensions of the reservoir. Thus, elongated
reservoir cross
 section favour homogeneity of supply motors.  For
this reason, the solutions of the
TASEP-LK$_m$ can provide a good approximation of the system when the cross
section
of the reservoir is large enough (see figure~\ref{finite_diff1}).
Hence,  we have shown that in some
geometries the reservoir is approximately homogeneous and we quantitatively
recover, as expected, the results previously presented for the TASEP-LK$_m$.

\section{Discussion}
We have introduced a theoretical framework to investigate the regulatory effects
of a finite closed reservoir of motor proteins on the formation of traffic jams
on
cytoskeletal filaments. 
In particular, we have studied two models with limited resources of
motors and their binding sites. 
 They represent finite processive motors
moving along
a filament in contact with a finite reservoir of motors, and describe how the
emergence
of traffic jams can be regulated by the supply-demand of molecular motors in
the
system. Our model can be seen as a generalization of exclusion processes with
Langmuir
kinetics and  infinite resources indicated as TASEP-LK$_\infty$ in the
text~\cite{Par03,parmeggiani2, evans_shock_2003}. In the present formulation we
study the impact on transport characteristics of a limited reservoir of
particles, as a possible experimental situation in \textit{in vitro} and
\textit{in vivo} experiments, recovering previous results in some limiting
cases. The competition for motor proteins is determined by the fraction $m$
between the total concentration of motor proteins and the concentration of
binding sites (tubulin dimers).   

The analytical solutions are confirmed by numerical simulations and they allow
us to characterize the dependence between motor crowding phenomena and supply,
thus leading to propose a novel determinant of traffic jams: the competition
parameter $m$.  Although  the  proposed theory can be far from
real \textit{in vivo} situations, it implies that in different regions  of the
cytoskeletal network the parameter $m$ can locally change
according to inhomogeneities in motor protein and filament concentrations or,
more
in general, availability of binding sites in competition with other proteins
binding to cytoskeletal  filaments.

Our model predicts that the onset of jams and other motor traffic features can
be
  determined by the local supply-demand constraint. 
For instance, depending on the local tubulin concentration, crowding phenomena
set up differently upon changes in $m$. The simple supply-demand balance might
then be  relevant for traffic regulation  inside the cell.

We have first analysed a system with a homogeneous distribution of particles in
the reservoir (TASEP-LK$_m$, Section~\ref{sec::MF}), for which we are able to
provide  analytical solutions. 
When the amount of motor proteins in the system is finite, there appears a
pronounced depletion of the motor reservoir at small values of  $m$
(figure~\ref{fig:finite1_c}). In this regime we have found strong
competition
of the binding sites for the relatively few motors  in the system.
We have shown that it is possible to devise quantitative relationships between
the  motor density on the filament and the total
availability of resources.  By measuring  the concentration of free
motors in the reservoir it is possible to extract information on the crowding of
the filament. Interestingly, when either the amount of binding sites or the
concentration of motors is large ($\kappa^\infty$ or $K^\infty$ tend to
infinity) we found a critical behaviour. For values of $m<m_c$ we observe the
very strong depletion of the reservoir for which most of the motors in the
system are bound to the
filament. Interestingly, in this regime, the concentration of motors in solution
remains finite although its fraction with respect to
the total concentration converges to zero.  
This is a subtle result since, even in the large $L$ limit
(i.e. large $\kappa^\infty$), motors still dissociate from the filament.
Accordingly,  one would
naively expect  a finite fraction of motors in the solution.

We then map  two different phase diagrams of the system: one as proposed
in~\cite{Varg2} with $\beta/p$ and $\rho_\ell^\infty$ as control parameters, the
other with the rescaled total concentrations of tubulins and motors
($\kappa^\infty$ and $K^\infty$ respectively). We have  investigated those
phase diagrams at different supply-demand ratios $m$ and
underlined the traffic effects introduced by the finite pool of motors. The
$(\rho_\ell^\infty,\beta/p)$ phase diagram is however difficult to
reconstitute experimentally, since the corresponding parameters cannot be
easily 
modified independently. Instead, the novel phase diagram introduced here allows
to study the effects of limited resources in a straightforward way.
By examining the phase diagrams it becomes apparent how the presence of a
shock on the filament is affected by the limitation of resources: at low
motor concentrations 
 or high amount of cytoskeletal binding sites, the spike phase LDs
dominates the $(\rho_\ell^\infty,\beta/p)$ phase diagram. An LD-HD phase is
always achievable, but it might only appear at the very low values of the
end-dissociation rate $\beta$. Moreover, from the ($\kappa^\infty$, $K^\infty$)
phase diagram it is evident that, starting from an LD-MC phase, the LDs phase
can only be reached after a passage to the LDn phase, while starting from the
LD-HD phase the LDn phase cannot be  found.  We find  that the phase diagram of
 TASEP-LK$_m$  interpolates between the antenna model neglecting
excluded volume interactions and  TASEP-LK$_{\infty}$, with a phase
diagram characterized by excluded volume interactions.  The TASEP-LK$_m$ model
thus quantifies how excluded volume determines active motor
protein transport.

The formalism developed for a single filament can be extended to more general
cases and in particular to investigate a population of
filaments, as  has been  done before for a multi-TASEP model with finite
resources~\cite{CianFinite}. Nevertheless, the results we obtain for
$\beta=0$ remain valid for any other type of motor dynamics on the segment, e.g.
bi-directional motion, and for any  configuration of filaments immersed in
a homogeneous reservoir, as in a network of cross-linked or branched filaments
mimicking a cytoskeleton, as studied in~\cite{NeriTT, NeriTTT}. 

In the  section~\ref{sec::diff} we have considered a reservoir of diffusive
particles in
order to investigate the impact of diffusion and  of concentration
gradients on the TASEP-LK$_m$ phenomenology.  With
finite diffusion the active transport indeed creates a stationary gradient of
particles in the reservoir. The presence of such a gradient introduces spatial
inhomogeneities in the motor-filament attachment rates.
  Numerical simulations
in section~\ref{sec::diff} revealed an important coincidence of phase behaviour
for these two distinct systems, namely (i) the TASEP-LK$_m$  and (ii) TASEP-LK
model with explicit motor diffusion.  In particular, upon changing $m$, both
domain wall and phase transition behaviour is qualitatively similar in both
systems.   Decoupling  supply-demand
effects and motor diffusion can therefore already provide  a
qualitative  understanding of the main  system features.

We stress that, although the two models are different, we can recover the
TASEP-LK$_m$ in certain limiting cases: in fact  we have shown by numerical
simulations, and explained with a theoretical argument, how the change of the
reservoir geometry results in very similar outputs of the two models. 
By increasing the cross section of the reservoir, we observe the
convergence of the motor density profile along the filament to the mean field
prediction of the TASEP-LK$_m$. Therefore we expect our analytical theory to
remain valid and provide quantitative good
results when the reservoir is sufficiently large (as in many \textit{in
vitro} systems).  

This work introduces an important parameter, the finite resources parameter $m$,
describing the competition between the concentration of polymerized cytoskeletal
filaments and the number of motors in solution, to the theoretical description
of \textit{in vitro} experiments as performed in~\cite{Varg2}. Those
experiments, indeed, have been done at very low concentration of tubulin (pM)
with respect to the concentration of motors (nM). As a consequence, the system
studied was in the infinite resource limit. 
With the TASEP-LK$_m$ it is now also possible to study regimes for which the
tubulin concentration is comparable to the motor protein concentration, and
thus the
competition between supply and demand of polymer binding sites for motors is
relevant. The phase diagram in the ($\kappa^\infty$,$K^\infty$)-plane opens the
possibility to tune directly the experimental setup (by changing the length of
microtubule filaments and/or the concentration of motor proteins) and
straightforwardly compare observed motor protein density profiles to our
analytical theory. 
Although we are aware that the models proposed only constitute an approximation
of \textit{in vivo} systems, we can speculate that the motor
density phenomenology presented
qualitatively holds when and where the competition for motor proteins becomes
relevant.  For instance we notice that our modeling framework is consistent with
\textit{in vivo} experimental results \cite{blasius_recycling_2013}. In this
work Blasius et al. have observed an exponential profile of kinesin-1 motors
along the length of CAD cell neurites and hippocampal axons. We could
qualitatively observe the same
behaviour for systems in LD phases and small $m$ (see e.g. black line in the top
left panel of Fig. 8), with characteristic lengths of the same order of
magnitude of the one found in~\cite{blasius_recycling_2013}. This is a promising
signature for future applications of our model, but a more quantitative and
careful investigation would require more information on the physiological
parameters and it is however beyond our presentation.

Through the  control of the crowding effects in the different phases, the
competition for resources can also  affect another important
biological process, namely the length-regulation of the filament. In fact it is
well
known that the motion of motor proteins (e.g. kinesin-8 Kip3p) at the filament
tip can regulate its depolymerization~\cite{VargaCell, klein_filament_2005,
Johann, Melbinger, Kuan}. In this perspective, it would be interesting to study
how the competition for resources regulates the filament depolymerization rate,
and to relate filament length regulation to the phase diagrams of TASEP-LK$_m$
that we have presented here. Note that to study the filament length regulation
at finite resources it is also necessary to consider the concentration of
tubulin dimers in the flow chamber solution/cytosol.  Such a study is out of the
scope of this paper.

\ack
We are grateful to P.~Greulich, C. Leduc and P. Malgaretti for interesting
discussions.  IN is grateful to M. Bock for a critical reading of the
manuscript. LC is supported by an EMBO long-term fellowship co-funded by the
European Commission (EMBOCOFUND2010, GA-2010-267146). We thank the ICSMB at the
University of Aberdeen and the  HPC@LR in Montpellier for the computational 
time support. J-CW acknowledges the support by the Laboratory of Excellence
Initiative (LabEx) NUMEV, OD by the Scientific Council of the University of
Montpellier 2.

\appendix

\section{TASEP-LK$_{\infty}$: density profiles and phase diagrams at infinite
resources}\label{App:A}
We consider here the TASEP-LK$_{\infty}$ model with  with infinite resources  
suc that the  the total motor concentration $c$
equals to the concentration of motors in the reservoir $c_u=c$.  We have as
well for the renormalized motor concentrations $K=K^{\infty}$ (with $K =
\tilde{\omega}_Ac_u/\omega_D$, and  $K^{\infty} =
\tilde{\omega}_Ac/\omega_D$).
The continuous density profile $\rho(x)$ for TASEP-LK$_{\infty}$
follows from the solution of the continuum mean field equations \cite{Par03,
parmeggiani2}:
\begin{eqnarray}
 \Omega^{-1}_D[2\rho(x)-1]\partial_x\rho(x)+ K[1-\rho(x)] - \rho(x)
= 0\;, \label{eq:rhoCont}
\end{eqnarray}
with boundary conditions $\rho(x=1) = 1-\beta$ and $\rho(x=0) = 0$, where here
and in the rest of this appendix we have posed, for the sake of simplicity and
without loss of generality, $p=1$. In the following we write down the analytical
expression for the density profile $\rho(x;K, \Omega_{\rm D}, \beta)$ by solving
equation (\ref{eq:rhoCont}), and present the corresponding phase diagram. We
closely follow the procedure presented in~\cite{Par03, parmeggiani2}.
 
\subsection{Density profile}
We write the density profile in terms of the ``renormalized''
profile
$\sigma\left(x\right)=\frac{K+1}{K-1}(2\rho(x)-1)-1$, where we have dropped 
the dependency on $K$, $\Omega_{\rm D}$ and $\beta$ for convenience. The
renormalized profile
$\sigma(x)$ can have three different forms determining the phase of the
system. The system presents a low density phase (LD) where the density profile
is continuous, a low density-high density coexistence phase (LD-HD) and a low
density-maximal current coexistence phase (LD-MC) where the profile develops a
shock at a position $x_w$ separating a low density region from a high density
region (or a maximal current region).
Here we make a further distinction among the LD phase, and we name LDn (no jam)
a low density phase with no presence of jams at the end (as in the case shown in
figure~\ref{fig:finite4} (A) for small $m$). Instead,  we name as LDs (spike)
the
low density phase with an occupation spike on the last site (as the profile with
the smallest $m$ shown in figure~\ref{fig:finite4} (B)).

The profile $\sigma(x)$ is thus given
by:
\begin{itemize} 
\item LD-phase:  $\sigma(x) = \sigma_L(x)$
\item LD-HD phase: $\sigma(x) = \left\{\begin{array}{ccc}\sigma_L(x)&&x<x_w \\
\sigma_R(x)&&x>x_w \end{array}\right. $
\item LD-MC phase: $\sigma(x) = \left\{\begin{array}{ccc}\sigma_L(x)&&x<x_w \\
\sigma_{MC}(x)&&x>x_w \end{array}\right. $
\end{itemize}
with $\sigma_L$ the left boundary solution (for which $\rho_\alpha(0)=0$),
$\sigma_R$ the right boundary solution (for which $\rho_\beta(1) = 1-\beta$) and
$\sigma_{MC}$ the saturated maximal current solution (for which
$\rho_{MC}(1)=1/2$). The point $x_w$ is defined by the equation
$\sigma_L(x_w)+\sigma_{R}(x_w) = -2$ or equivalently $\rho_\alpha(x_w) =
1-\rho_\beta(x_w)$.   The left and right boundary solutions are given in terms
of
the real valued branches $W_{0}$ and $W_{-1}$ of the Lambert-W
function~\cite{corless_lambert_1996}.  We have for the expression of $\sigma_L$
\begin{eqnarray}
\fl  \sigma_{L}\left(x\right) =\left\{\begin{array}{ccc}
W_0\left(\frac{2K}{1-K}\exp\left(\Omega_{\rm D}\frac{(K+1)^2}{K-1}x
-\frac{2K}{K-1} \right)\right)&& K<1 \\
W_{-1}\left(\frac{2K}{1-K}\exp\left(\Omega_{\rm D}\frac{(K+1)^2}{K-1}x
-\frac{2K}{K-1}\right)\right)&& K>1\end{array} \right., \label{eq:sigmabla}
\end{eqnarray}
and  the right boundary solution $\sigma_R$ is given by
\begin{eqnarray}
\fl \sigma_{R}(x) = \left\{\begin{array}{ccc}
W_{-1}\left(\left(\frac{K+1}{K-1}(1-2\beta)-1\right)\exp\left(\Omega_{\rm
D}\frac{
(K+1)^2 } { K-1 } (x-1)
+ \left(\frac{K+1}{K-1}(1-2\beta)-1\right)\right)\right)&& K<1
\\W_0\left(\left(\frac{K+1}{K-1}(1-2\beta)-1\right)\exp\left(\Omega_{\rm
D}\frac{
(K+1)^2 } { K-1 } (x-1)
+ \left(\frac{K+1}{K-1}(1-2\beta)-1\right)\right)\right)
&& K>1\end{array} \right. .
\nonumber \\\label{eq:sigmablaT}
\end{eqnarray}

\begin{figure}[t]
 \begin{center}
 \hbox{
 \includegraphics[width =0.36\textwidth]{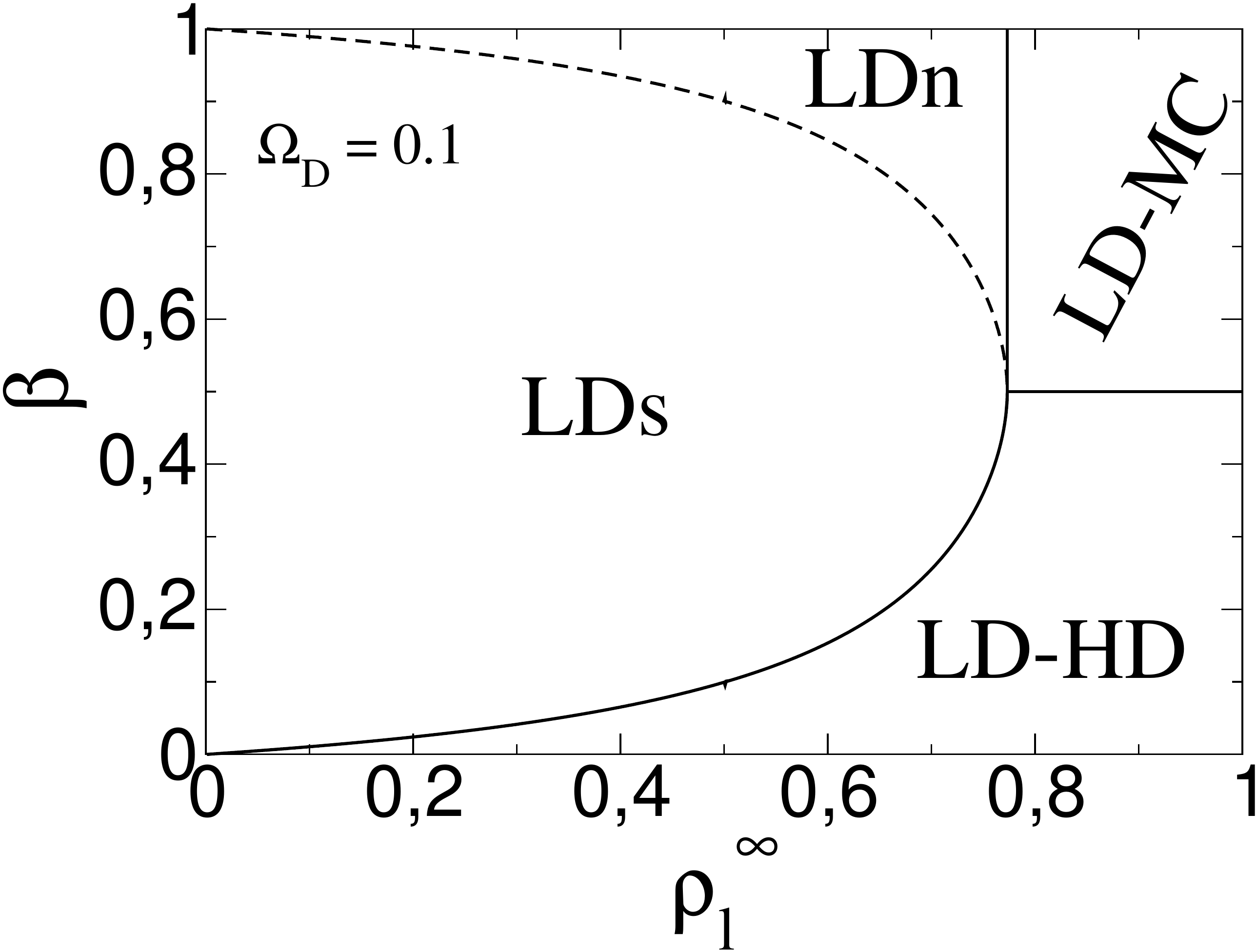}
 \includegraphics[width =0.3\textwidth]{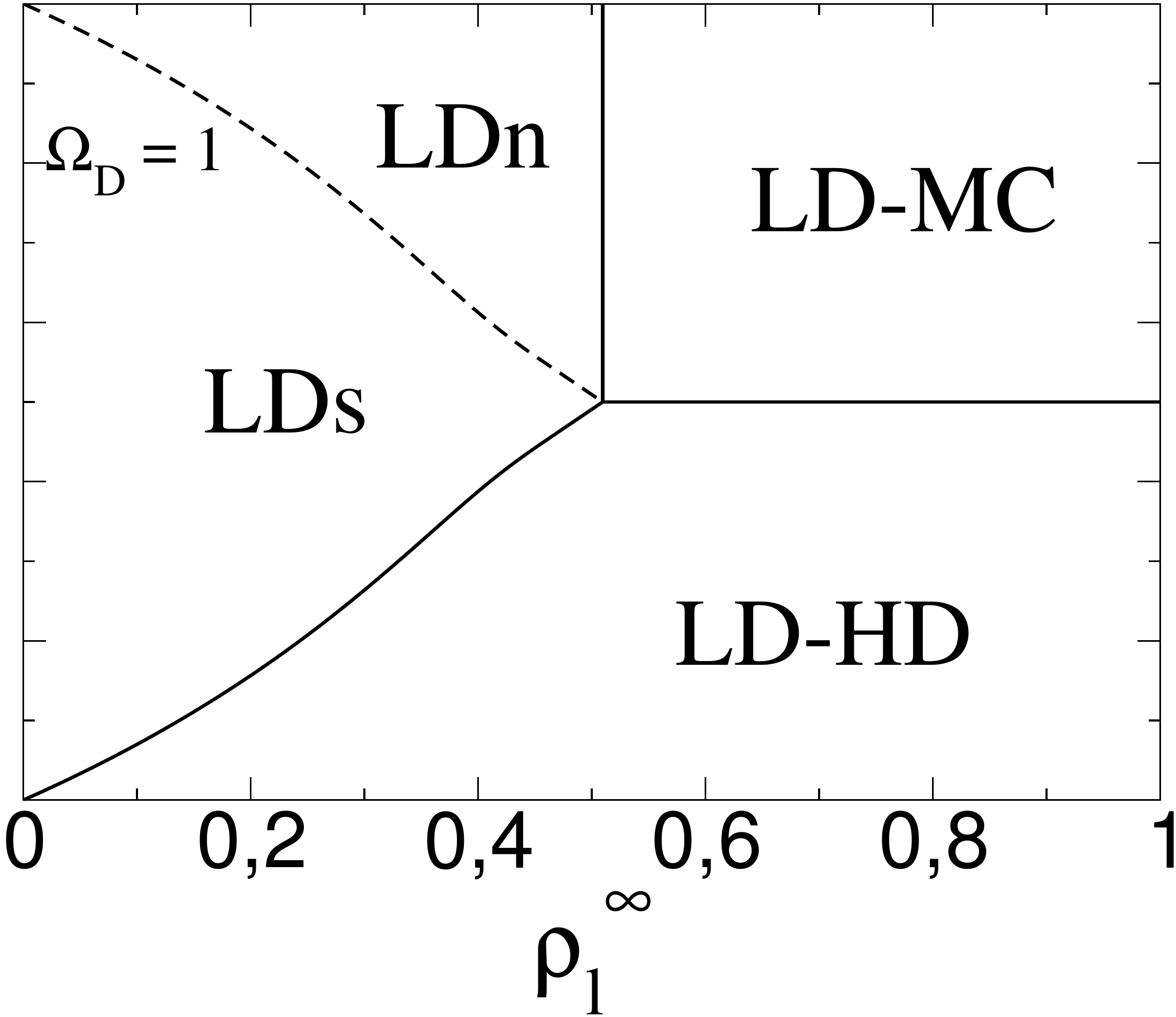}
 \includegraphics[width =0.3\textwidth]{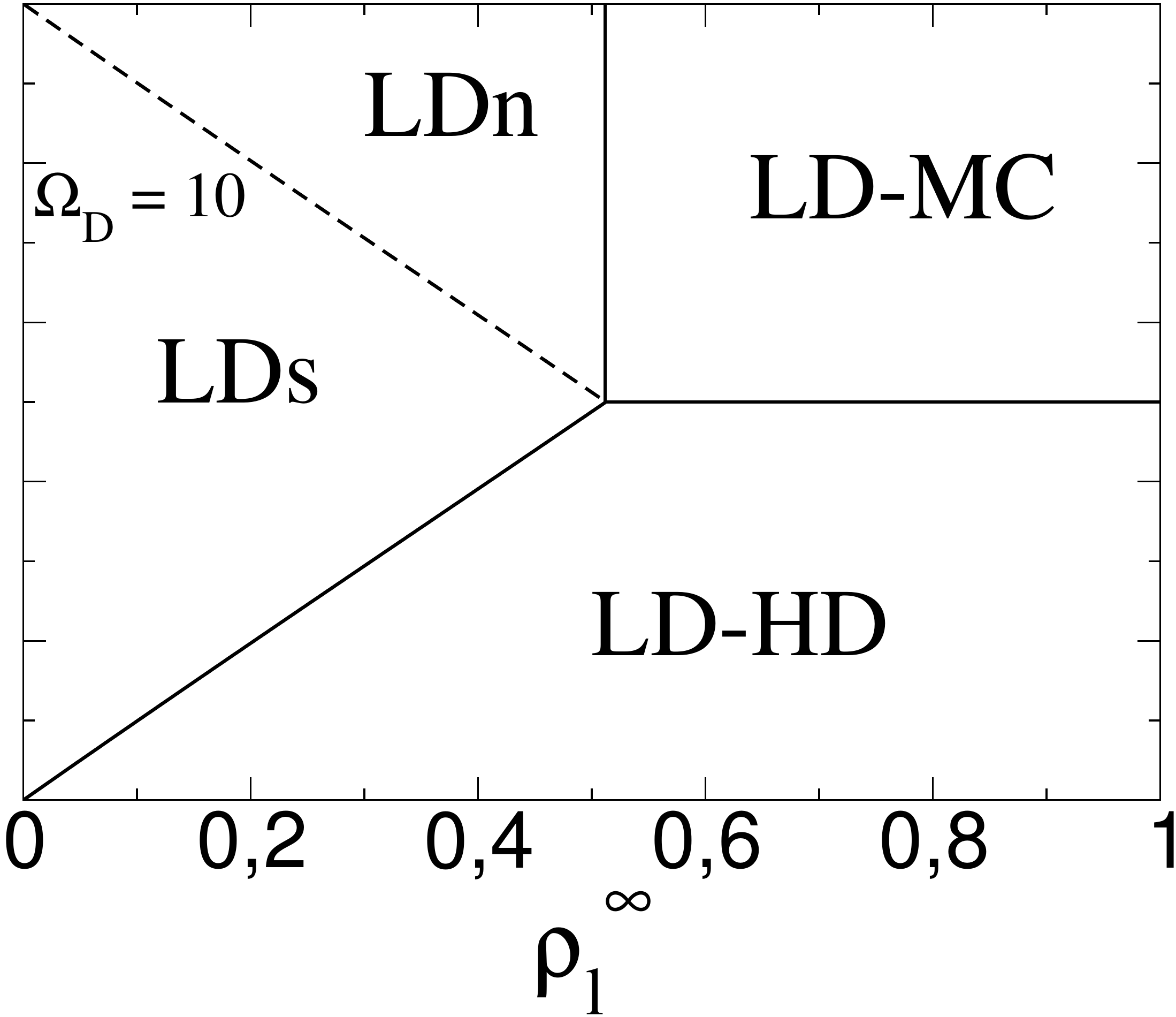}
 }
 \caption{The $(\rho_\ell^\infty, \beta)$-phase diagrams for
 TASEP-LK$_{\infty}$ for given values of $\Omega_{\rm D}$.
  }\label{fig:appendixLK}
 \end{center}
 \end{figure}
\subsection{Phase diagram for infinite resources}\label{app:infRes}
We now determine the $(\rho^{\infty}_{\ell}, \beta)$-phase diagram at infinite
resources for fixed values of $\Omega_{\rm D}$. In this case we have $K =
K^\infty$ and therefore $\rho^{\infty}_{\ell}=K/(K+1)$.
We get the following equations for the transitions between the LDs, LDn, LD-HD
and LD-MC phases:
\begin{itemize}
\item {\it LD-HD to LD-MC}: the transition happens at $\beta=1/2$ for
$K>1$.
 \item {\it LD-HD to LDs}: this transition occurs for $\beta<1/2$.  The
transition is  given by the
condition $\rho_\alpha(x=1) = 1-\rho_\beta(x=1) = \beta$.   Note that this
condition
corresponds to  $\rho_\alpha(x=1)(1-\rho_\alpha(x=1)) = \beta(1-\beta)$, or,
setting  the current to the left of the shock equal to the current leaving the
segment. From the
solution for $\rho_\alpha$, which follows from equation (\ref{eq:sigmabla}), we
find, for the critical $\beta<1/2$: 
\begin{eqnarray}
 \fl \beta =\left\{\begin{array}{ccc}
\left(\frac{K-1}{2(K+1)}\right)W_0\left(\left|\frac{2K}{K-1}\right|\exp\left(
\Omega_{\rm D}\frac{(K+1)^2}{K-1}
-\frac{2K}{K-1} \right)\right)  + \frac{K}{K+1}&& K<1\;, \\
\left(\frac{K-1}{2(K+1)}\right)W_{-1}\left(-\left|\frac{2K}{K-1}\right|\exp\left
(\Omega_{\rm D}\frac{(K+1)^2}{K-1}
-\frac{2K}{K-1}\right)\right)  + \frac{K}{K+1}&&  K>1\;. \end{array}
\right.\nonumber \\\label{eq:betacx}
\end{eqnarray}
\item {\it LDn to LD-MC}: this transition happens at the critical $K_c>1$ at
which $\beta_c=1/2$ in equation (\ref{eq:betacx}). Hence, we get the equation
\begin{eqnarray}
\fl \left\{\begin{array}{ccc} 
W_0\left(\left|\frac{2K}{K-1}\right|\exp\left(\Omega_{\rm D}\frac{(K+1)^2}{K-1}
-\frac{2K}{K-1} \right)\right)= -1&&  K\leq 1\;,  \\
W_{-1}\left(-\left|\frac{2K}{K-1}\right|\exp\left(\Omega_{\rm
D}\frac{(K+1)^2}{K-1}
-\frac{2K}{K-1}\right)\right) = -1&&  K\geq 1\;,
\end{array}\right.\label{eq:KCalc}
\end{eqnarray}
which provide the transition lines for values of $\beta>1/2$.
We can now use that $W_{0}(-1/e) = W_{-1}(-1/e) = -1$, to obtain the
result
\begin{eqnarray}
\begin{array}{ccccccc}
 \Omega_{\rm D} &=& &\frac{1}{K+1} +
\frac{(K-1)}{(K+1)^2}\ln\left(\frac{K-1}{2K}\right)&  && \Omega_{\rm D}<0.5\;, 
\\
K&=&1&& && \Omega_{\rm D} >0.5\;.
\end{array}
\label{eq:KcFunctionOmegaD}
\end{eqnarray}
\item {\it LDs to LDn}: this transition is given by the condition
$\rho_\alpha(x=1) = \rho_{i=L}$. Since, $\beta \rho_{i=L} =
\rho_\alpha(x=1)(1-\rho_\alpha(x=1))$ we get the condition $\rho_\alpha(x=1) =
1-\beta$. We thus see that when $\beta_c$ is the transition from LD-HD to LDs,
then $1-\beta_c$ denotes the transition from LDs to LDn.
\end{itemize}
In figure \ref{fig:appendixLK} we have drawn the corresponding
$(\rho^{\infty}_{\ell}, \beta)$-phase diagrams for the TASEP-LK$_{\infty}$.

\section{TASEP-LK$_m$: Phase transition lines for finite
resources}\label{sec:phaseTransFinite}
To get the phase diagrams for the TASEP-LK$_m$ at finite resources we are led to
substitute the concentration of motors at finite resources $K(\beta, \Omega_{\rm
D}, \kappa^\infty, K^\infty)$, (solution of the equations
(\ref{eq:currentConsx}), (\ref{eq:densityProfile}) and (\ref{eq:K1})) into the
equations for the phase transitions at infinite resources (found in
\ref{app:infRes}).

Below we present the analytical results obtained for the phase transition lines:
\begin{itemize}
 \item {\it LD-HD to LD-MC}: the critical transition line is given by:
\begin{eqnarray}
 \beta &=& 1/2  \ \  {\rm for} \ \ K(\beta, \Omega_{\rm D}, \kappa^\infty,
K^\infty)>1 \;.
\end{eqnarray}

\item {\it LD-HD to LDs}:
The critical value $\beta_c$ follows from solving an implicit equation that we
find from the substitution of $K(\beta, \Omega_{\rm D}, \kappa^\infty,
K^\infty)$
into the equation (\ref{eq:betacx}).
For instance, when $\Omega_{\rm D}\rightarrow\infty$, from
Eq.~(\ref{eq:currentConsx}) the density on the lattice corresponds to the one
that would be obtained in a closed system: $\rho=\rho_\ell=K/(K+1)$. From that
we get the critical beta $\beta_c=K/(K+1)$.

\item {\it  LDn  to LD-MC}:
We substitute the stationary value of $K(\beta=1/2, \Omega_{\rm D},
\kappa^\infty, K^\infty) = K^\ast$, which follows from setting $\rho(1)=1/2$ in
equation (\ref{eq:K1}), into the equation (\ref{eq:KcFunctionOmegaD}).   We then
have, for $\beta>1/2$, the solution
\begin{eqnarray}
 \begin{array}{ccccc}
 \Omega_{\rm D} &=& \frac{1}{ K^\ast+1} +
\frac{( K^\ast-1)}{( K^\ast+1)^2}\ln\left(\frac{ K^\ast-1}{2 K^\ast}\right) &&
\Omega_{\rm D}<0.5\;,  \\
K^\ast &=& 1 &&  \Omega_{\rm D}>0.5\;.
\end{array} \label{eq:transLDMC}
\end{eqnarray}

\item {\it LDs to LDn}:   If $\beta_c$ denotes the LDs to LD-HD transition, then
$1-\beta_c$ denotes again the transition from LDs to LDn. Indeed, the transition
from LDs to LD-HD is denoted by the condition $\rho_\alpha(x=1; K,\Omega_{\rm
D}) = \beta$ and the transition from LDs to LDn by $\rho_\alpha(x=1; K,
\Omega_{\rm D}) = 1-\beta$.  For finite resources, $K\neq K^\infty$ is a
function of $\beta$ itself.  However, the dependency of $K$ on $\beta$ is given
by the current at the end tip $\rho(x=1)(1-\rho(x=1)) = \beta(1-\beta)$, which
is independent of the transformation $\beta\rightarrow 1-\beta$.  Hence, also
for finite resources, the critical transition $\beta_c$ from LDs to LDn  is
given by $1-\beta_c$ the transition of LDs to LD-HD.
\end{itemize}

Let us finally present some formal analytical result on the nature of the phase
diagrams in figure \ref{fig:finite2}. We see that when $m$ becomes smaller the
LD-MC phase and LD-HD phase reduce in size and the LDs phase becomes more
prominent. While the LD-HD phase will only disappear for $m\rightarrow 0$, the
LD-MC phase will disappear for some $m=m^\ast>0$. To find this ``critical''
$m^\ast$  we set
$\rho_\ell=1$ in equation (\ref{eq:transLDMC}) to find for $\Omega_{\rm D}>1/2$,
or equivalently $m\in[1/4,1/2]$:
\begin{eqnarray}
  m^\ast &=& \frac{1}{2}\left(1-\frac{1}{4\Omega_{\rm D}}\right)\;,
\end{eqnarray}
while for $\Omega_{\rm D}<1/2$ and thus $m<1/4$ we need to solve
\begin{eqnarray}
 \fl \frac{1+4\Omega_{\rm D}}{4(1-m^\ast)} = 1 +
\frac{4(2m^\ast-1)\Omega_{\rm D}+1}{1+4\Omega_{\rm
D}}\ln\left(\frac{4(2m^\ast-1)\Omega_{\rm D}+1}
{
2(4m^\ast\Omega_{\rm D}+1) } \right) \;.
\end{eqnarray}

\section{TASEP-LK: Reservoir with excluded volume}\label{App:B}
Let us  consider a reservoir in which particles feel in the reservoir an
exclusion
interaction.  A formula for the bound density has then been derived in
\cite{Klumppxx} (compare equation (3) in \cite{Klumppxx} with equations (4) and
(6) in this work). We now show
 the connection between these equations.   

Our results can be easily generalized to the case of a finite
reservoir with exclusion interactions.   To define such a reservoir, we have to
consider particles with a
finite volume $v_m$, leading to the
fraction $\phi_u = c_uv_m$ of the volume being
occupied.   Hereby we have assumed that the filament itself has no volume
(but this can be easily corrected when renormalizing the volume). 

The equation for the mass conservation (\ref{eq:NConsx}) and the hydrodynamic
equation for the bound
density (\ref{eq:currentConsx}) remain the same. What does change with volume
 exclusion  is  the balance between
reservoir and segment.  In fact, the detachment current will now be $J_D =
\omega_{\rm D}\rho(1-\phi_u)$, and the exit current $J_{\rm end} =
\rho(1)[1-\rho(1)]=\beta(1-\phi_u)\rho_{i=L}$. We recover the result without
exclusion $J_D = \omega_{\rm D} \rho$ when $v_m\rightarrow 0$ or equivalently
$\phi_u \rightarrow 0$. The balance equation becomes then 
\begin{eqnarray}
\rho = \frac{K - \Omega^{-1}_D[1-\rho(1)]\rho(1)}{K + (1-\phi_u)}.
\end{eqnarray}
It is useful to write $\phi_u = K\left(v_m\omega_{\rm
D}/\tilde{\omega}_A\right)$. 
Accordingly $K$ is again  function of $\rho(1)$:
\begin{eqnarray}
\fl  2\left(1-v_m\omega_{\rm D}/\tilde{\omega_{\rm A}}\right)K=-1 -Q
\nonumber \\
+\sqrt{1+
Q^2
+ 4 \: \left(1-v_m\omega_{\rm D}/\tilde{\omega_{\rm A}}\right)\left\{K^{\infty}+
\kappa^\infty[1-\rho(1)]\rho(1)/\Omega_{\rm D}\right\}
}, \nonumber \\\label{eq:K1x}
\end{eqnarray} 
with 
\begin{eqnarray}
 Q = \kappa^{\infty}-\left(1-v_m\omega_{\rm D}/\tilde{\omega_{\rm
A}}\right)K^{\infty}.
\end{eqnarray}

It is interesting to notice that in the case $\beta=0$ we recover the formula
(3) present in \cite{Klumppxx} when setting $v_m=1$, such that $\phi_u = c_u$,
$\rho(1)=1$,  $V = N_{\rm chan}L$, $K^\infty =
\frac{\pi_{\rm ad}}{\epsilon} \frac{N_{\rm mo}}{N_{\rm chan}L}$, $\kappa^\infty
= \frac{\pi_{\rm ad}}{\epsilon} \frac{1}{N_{\rm chan}}$,
$\omega_{\rm D}/{\tilde\omega_{\rm A}}  = \epsilon/\pi_{\rm ad}$ and $K =
\pi_{\rm
ad}/\epsilon c_u$. For the definitions of the quantities see~\cite{Klumppxx}.

\section{Numerical simulations}\label{appendix_sim}
\subsection{Unstructured reservoir}%
The results presented in figures~\ref{fig:finite1_c} and \ref{fig:finite4} have
been obtained by simulations with a continuous time Monte Carlo based on the
Gillespie algorithm~\cite{gill1,gill2}, also known as the Kinetic Monte Carlo
algorithm~\cite{bortz_new_1975}.

The reservoir of the TASEP-LK$_{m}$ is unstructured, in the sense that it is
represented by a variable $N_u$
 that counts the number of available particles. When a particle detaches (binds)
the lattice, $N_b$ is increased
 (decreased) by one. Since in the model with finite resources the attachment
rates depends on the number of available
 particles, the microscopic rate $\omega_{\rm A} = \tilde\omega_{\rm A} c_u = 
\tilde\omega_{\rm A} N_u/V$ needs
 to be updated when a binding or unbinding event occurs.

\subsection{Finite diffusion of particles in the reservoir}
\begin{figure}[!h]
 \centerline {
 \includegraphics[angle=0,width=11cm]{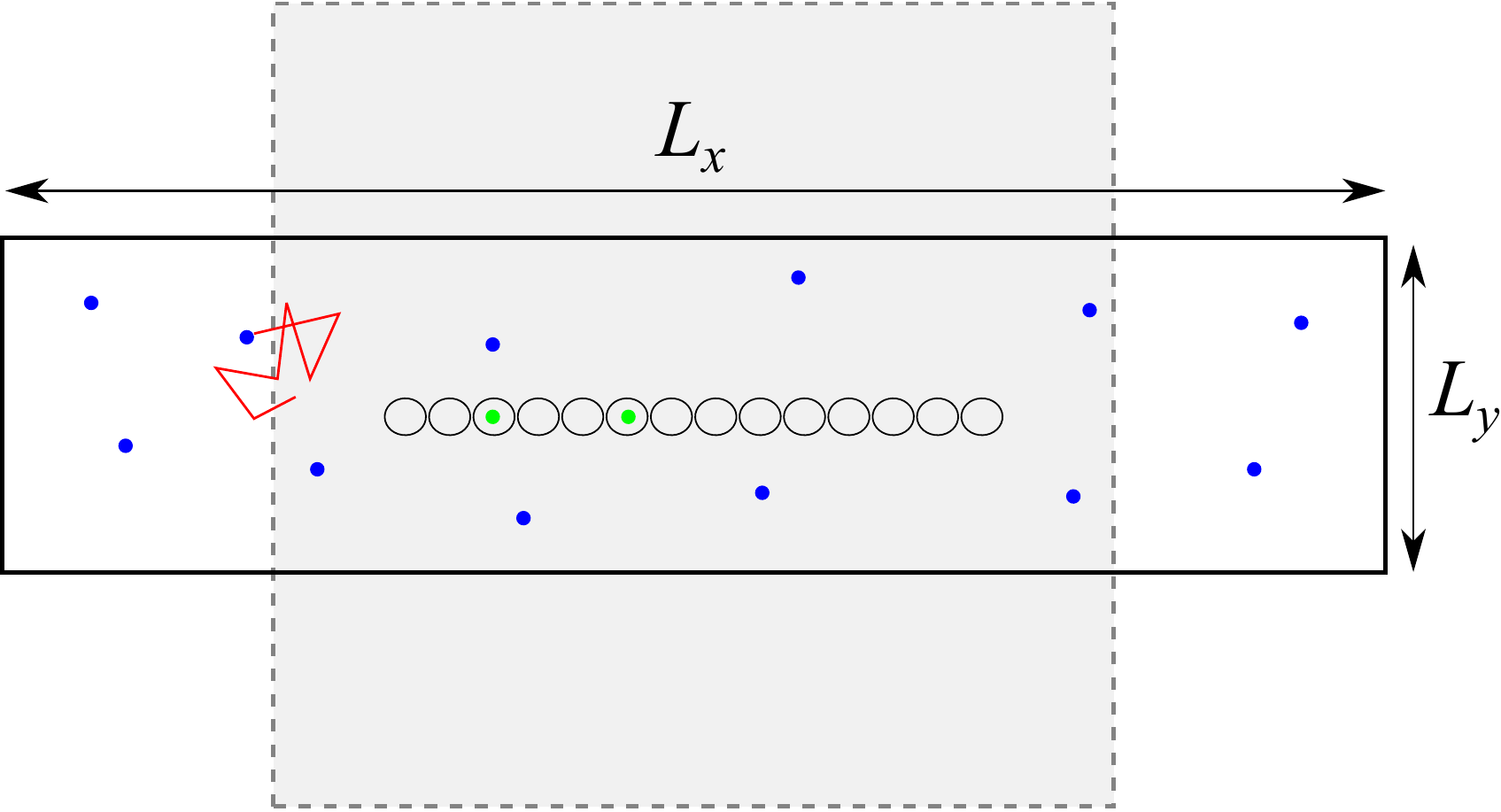}
 }
 \caption{The reservoir is a rectangular surface. The particles in the
 reservoir
 (blue dots) are diffusing with a diffusion
  coefficient $D$, without interactions (a typical trajectory is shown in red).
 The sites of the TASEP-LK filament are modeled by $L$
 disks (reaction surfaces) of diameter $a$. Particles can attach (green dots)
 or
 detach from these disks.
  In order to check the effect of the cross-section $L_y$, while keeping the
 size of the reservoir constant, we change $L_x$ and $L_y$ such that the volume
 of the reservoir
 $L_x\times L_y$ remains constant. This is illustrated with the gray rectangle
 which has a larger cross-section than the ``white`` rectangle.
 }
 \label{finite_diff2}
 \end{figure}

Figures~\ref{fig::diff0} and \ref{finite_diff1} have  been produced by
considering a spatially extended reservoir with particle
 diffusion. For these simulations, the TASEP-LK filament is immersed in a
spatial reservoir with diffusive particles as defined in figure
\ref{fig::models}.
 The protocol used to obtain the different curves in figure
\ref{finite_diff1} is explained in
figure~\ref{finite_diff2}, in particular 
how we have varied the size of the reservoir.

For simplicity, the reservoir is two-dimensional with a rectangular shape. 
The lengths of the edges of the rectangle are defined by $L_x$ and $L_y$
 (see figure~\ref{finite_diff2}). The filament has a length $L_f$ corresponding
to $L$ disks with a reaction surface of diameter $a\approx L_f/L$. Since the
dimensions of the
filament are constant during our simulations, we set the length unit $a$ of the
system equal to the
diameter of a disk: $a=10^{-8}m$  corresponds to the typical order of
magnitude of a step done by motor proteins of the kinesin
family~\cite{HowardBook}.
In the simulations shown in figure~\ref{finite_diff1}, we choose $N=10^3$
particles and $L=10^3$ sites
so that $m=1$. The reservoir has a constant surface $L_x\, L_y=50\mu m^2$,
and so the concentration is kept constant ($c=3.32\cdot10^{-11}mol\,m^{-2}$ for
all curves in figure~\ref{finite_diff1}, with a length of the filament of
$L_f=10\mu m$ and $\Omega^\infty_{\rm
A}=1$). The other parameters $\beta/p$ 
and $\Omega_{\rm D}$ determine the phase of the system and vary from graph to
graph.\\
 
The TASEP-LK filament is thus modeled as a sequence of disks playing the role of
a reaction volume: if the site in not occupied,
 then a diffusive particle in the reaction volume can attach to this site at a
constant rate $k$. It implies that
the rate of attaching to the site $i$ is $\omega_{\rm A}(i)=k\,
n(i)=\tilde\omega_{\rm A} c_u(i)$ where $n(i)$ and $c_u(i)$ are respectively the
number of particles and the concentration of
 the reaction volume associated to the site $i$. Thus we have
 $\tilde\omega_{\rm A}=k\, v_R$ where $v_R$ is the volume of reaction (here
the disk area). The other update rules on the
 filament are unchanged: particles can perform a discrete jump to the right
with a rate $p$, which sets the timescale of the system; if a particle detaches
at a
 rate $\omega_{\rm D}$ or $\beta$, then it performs diffusion in the reservoir.
We fix $p=50s^{-1}$ to the order fo magnitude of measured rates for kinesins
~\cite{schnitzer_kinesin_1997}. 
The outcomes of simulations, figure~\ref{finite_diff1}, show the effect of
changing the reservoir size, namely by increasing $L_y$. At the same time, we
decrease $L_x$
such that the area of the reservoir and hence the other parameters remain
constant, like the total concentration $c=N/(L_xL_y)$ (and so
$\rho^\infty_\ell$)
and the parameter $m=N/L$.
 In that way, the system remains at the same point of the phase diagram, as we
can see by considering e.g.
 figure~\ref{fig:finite2}.

The particles in the reservoir evolve according to a Brownian dynamics defined
by equations
(\ref{BDEq}-\ref{BDEq3}) in subsection \ref{sub::mod_diff}.  
We introduce now the finite difference $\Delta t$ instead of the derivatives in
the equation of motion. The $\delta(t-t')$ in the amplitudes of the white noise
is replaced by $1/\Delta t$ where $\Delta t=t-t'$~:
\begin{eqnarray*}
\frac{\vec r(t+\Delta t)-\vec r(t)}{\Delta t}&=&\vec\xi(t)\,,\\
r_a(t+\Delta t)&=&r_a(t)+\sqrt{2D\Delta t}\eta_a\,,
\label{BDEqFD}
\end{eqnarray*} 
where $\eta_a$ is a random number generated with a gaussian distribution with
unit variance.

After one update of the filament, the diffusion equation for the diffusing
particles in the reservoir are
integrated during the time of evolution of the filament given by the time
interval $\tau=S_R^{-1}\ln(1/r)$~\cite{gill1,gill2},
 where $S_R$ is the sum of all the possible rates and $r$ is a random number
uniformly distributed in $(0,1]$. If the position
of a particle is updated outside the box, the move is rejected. To reduce these
boundary effects, we split this global interval of integration in small
enough time intervals.

 \section*{References}
 \bibliographystyle{ieeetr}
 \bibliography{bibliography}

\end{document}